\documentclass{emulateapj}
\usepackage{hyperref}
\usepackage{amssymb}
\usepackage{pifont}

\usepackage{amsmath}
\usepackage{epsfig}
\usepackage{natbib}
\usepackage{color}
\usepackage{epstopdf}
\usepackage{xspace}

\usepackage{hyperref}
\usepackage{nicefrac}
\hypersetup{
    colorlinks,
    citecolor=black,
    filecolor=blue,
    linkcolor=blue,
    urlcolor=blue
}
\usepackage{listings}
\usepackage{color}

\definecolor{dkgreen}{rgb}{0,0.6,0}
\definecolor{gray}{rgb}{0.5,0.5,0.5}
\definecolor{mauve}{rgb}{0.58,0,0.82}

\newcommand{\cmark}{\ding{51}}%

\newcommand{\xmark}{\text{\ding{55}}}

\def \kms {km\ s$^{-1}$\xspace}
\def \Msol {$\mathrm{M_\sun}$\xspace}  

\def \hMpc {$h^{-1}$\ Mpc\xspace}
\def \hpc {$h^{-1}$\ pc\xspace}
\def \Mvir {$M_{\mathrm{vir}}$\xspace}

\def \Rvir {$R_{\mathrm{vir}}$\xspace}
\def \Rs {$R_{\mathrm{s}}$\xspace}
\def \Vmax {$V_{\mathrm{max}}$\xspace}

\def \Rmax {$R_{\mathrm{max}}$\xspace}
\def \Vpeak {$V_{\mathrm{peak}}$\xspace}
\def \Mpeak {$M_{\mathrm{peak}}$\xspace}
\def \Msub {$M_{\mathrm{sub}}$\xspace}
\def \Mvirzero {$M_{\mathrm{vir}}(z = 0)$\xspace}
\def \rockstar {{\sc{rockstar}}\xspace}
\def \unboundthreshold {\texttt{unbound\_threshold}\xspace}

\shorttitle{Caterpillar Project}
\shortauthors{Griffen et al.}

\begin{document}
\title{The $Caterpillar$ Project: A Large Suite of Milky Way Sized Halos}
\author{Brendan F. Griffen\altaffilmark{1}, Alexander P. Ji\altaffilmark{1}, Gregory A. Dooley\altaffilmark{1},  Facundo A. G\'omez,\altaffilmark{2}, Mark Vogelsberger\altaffilmark{1}, Brian W. O'Shea\altaffilmark{3,4}, Anna Frebel\altaffilmark{1}}

\altaffiltext{1}{Kavli Institute for Astrophysics and Space Research
  and Department of Physics, Massachusetts Institute of Technology, 77
  Massachusetts Avenue, Cambridge, MA 02139, USA; \href{mailto:brendan.f.griffen@gmail.com}{brendan.f.griffen@gmail.com}}
\altaffiltext{2}{Max-Planck-Institute for Astrophysics, Karl-Schwarzschild-Str. 1, 85740 Garching, Germany}
\altaffiltext{3}{Department of Physics and Astronomy,  Department of Computational Mathematics, Science and Engineering, and National Superconducting Cyclotron Laboratory, Michigan State University, 567 Wilson Road, East Lansing MI 48824, USA}
\altaffiltext{4}{Joint Institute for Nuclear Astrophysics - Center for Evolution of the Elements, East Lansing, MI 48824}

\begin{abstract}
We present the largest number of Milky Way sized dark matter halos simulated at very high mass ($\sim$\ $10^4$ \Msol/particle) and temporal resolution (5 Myrs/snapshot) done to date, quadrupling what is currently available in the literature. This initial suite consists of the first 24 halos of the \href{http://www.caterpillarproject.org}{\textit{Caterpillar Project}}\footnote{Project Website: \url{http://www.caterpillarproject.org}} whose project goal of 60 -- 70 halos will be made public when complete. We do not bias our halo selection by the size of the Lagrangian volume. We resolve $\sim$20,000 gravitationally bound subhalos within the virial radius of each host halo. Improvements were made upon current state-of-the-art halo finders to better identify substructure at such high resolutions, and on average we recover $\sim$4 subhalos in each host halo above 10$^8$\ \Msol which would have otherwise not been found. The density profiles of relaxed host halos are reasonably fit by Einasto profiles ($\alpha$ = 0.169 $\pm$ 0.023) with dependence on the assembly history of a given halo. Averaging over all halos, the substructure mass fraction is $f_{m,subs} = 0.121 \pm 0.041$, and mass function slope is d$N$/d$M\propto M^{-1.88 \pm 0.10}$. We find concentration-dependent scatter in the normalizations at fixed halo mass. Our detailed contamination study of 264 low-resolution halos has resulted in unprecedentedly large high-resolution regions around our host halos for our fiducial resolution (sphere of radius $\sim1.4 \pm 0.4$ Mpc). This suite will allow detailed studies of low mass dwarf galaxies out to large galactocentric radii and the very first stellar systems at high redshift ($z$ $\geq$ 15).

\end{abstract}

\keywords{galaxy: halo -- galaxy: formation --- cosmology: theory}

\section{Introduction}
\label{sec:introduction}

Under the current paradigm of structure formation
\citep{White:1978uk} stellar halos of large
galaxies such as the Milky Way are believed to be primarily formed as
a result of the accumulation of tidal debris associated with ancient
as well as recent and ongoing accretion events
(\citealt{Helmi:2008dg}, \citealt{Pillepich:2015bya}).  In principle,
the entire merger and star formation history of our Galaxy and its
satellites can be probed with their stellar contents (i.e., the ``fossil
record''; \citealt{Freeman:2002kz}) because this information is not
only encoded in the dynamical distribution of the different
Galactic components, but also in the stellar chemical abundance
patterns (e.g., \citealt{Font:2006hx,Gomez:2010dx}). 

To further map out the structure and composition of the various
components of the Milky Way, large scale observational efforts are now
underway. Several surveys such as {\sc{rave}}
\citep{Steinmetz:2006cf}, {\sc{segue}} \citep{Yanny:2009fo}, {\sc{apogee}}
\citep{Majewski:2010bi}, {\sc{lamost}} \citep{Deng:2012ei} and {\sc{galah}} \citep{Freeman:2012wf} have collected medium-resolution
  spectroscopic data on some four million stars primarily in the
  Galactic disk and stellar halo. There are also ongoing
  large-scale photometric surveys such as Pan-STARRS \citep{Kaiser:2010gr} and
  SkyMapper Southern Sky Survey \citep{Keller:2013ff} mapping nearly the
  entire sky. Soon, the {\sc{gaia}} satellite \citep{Perryman:2001cp}
  will provide precise photometry and astrometry for another one
  billion stars.

Studies of individual metal-poor halo stars have long been used to
establish properties of the Galactic halo, such as the metallicity
distribution function, to learn about its history and evolution. More
recently, the discoveries of the ultra-faint dwarf galaxies (with
$L_{\rm tot} \le 10^{5}\,L_{\odot}$) in the northern Sloan Digital Sky
Survey (SDSS) and the southern Dark Energy Survey (DES) have shown
them to be extremely metal-deficient systems which lack metal-rich
stars with ${\rm [Fe/H]\gtrsim-1.0}$. To some extent they can be
considered counterparts to the most metal-poor halo stars. They extend
the metallicity-luminosity relationship of the classical dwarf
spheroidal galaxies down to $L_{\rm tot} \sim 10^{3}\,L_{\odot}$
\citep{Kirby:2008fs}, and due to their relatively simple nature, they
retain signatures of the earliest stages of chemical enrichment in
their stellar population(s). Indeed, the chemical abundances of
individual stars in the faintest galaxies suggest a close connection
to metal-poor halo stars in the Galaxy \citep{Frebel:2015kk}.

This comes at a time when there is still uncertainty over what role
dwarf galaxies play in the assembly of old stellar halos because the
true nature of the building blocks of large galaxies (e.g.,
\citealt{Helmi:2000go}, \citealt{Johnston:2008jp},
\citealt{Gomez:2010dx}) are not yet fully understood. Nevertheless,
observations of the, e.g., the Segue~1 ultra-faint dwarf suggest that these
faintest satellites could be some of the the universe's first galaxies
(presumably the building blocks) that survived until today
\citep{Frebel:2012ja, Frebel:2014gi}. They would thus be responsible for
the Milky Way's oldest and most metal-poor stars. 

This wealth of observational results offers unique opportunities to
study galaxy assembly and evolution and will thus strongly inform our
understanding of the formation of the Milky Way. Along with it, the
current dark energy plus cold dark matter paradigm ($\Lambda$CDM) can
be tested at the scales of the Milky Way and within the Local Group.
But to fully unravel the Galaxy's past and properties, theoretical and
statistical tools need to be in place to make efficient use of data.

For over three decades now, numerical simulations of structure
formation have consistently increased in precision and physical
realism (see \citealt{Somerville:2014una} for a review). Originally,
they began as a way to study the evolution of simple \textit{N}-body
systems (e.g., merging galaxies; \citealt{Aarseth:1963tc},
\citealt{Toomre:1972ji}, \citealt{White:1978ub} and globular clusters;
\citealt{Henon:1961tz}) but with the advent of better processing power
and more sophisticated codes (e.g., \citealt{Springel:2010hx},
\citealt{Hopkins:2015tz}, \citealt{Bryan:2014ev}), \textit{N}-body solvers are now fully
coupled to hydrodynamic solvers allowing for a comprehensive treatment
of the evolution of the visible Universe
(e.g., \citealt{Vogelsberger:2014gwa}, \citealt{Schaye:2015gk}).

The most efficient method of studying volumes comparable to the Local
Group whilst maintaining accurate large scale, low-frequency
cosmological modes is via the \textit{zoom-in} technique
(\citealt{Katz:1994wv}, \citealt{Navarro:1994uw}). This technique
allows one to efficiently model a limited volume of the Universe at an
extremely high resolution. Owing to the extreme dynamic range offered
by such simulations, both the inside of extremely low mass,
gravitationally bound satellite systems can be studied along side the
hierarchical assembly of their host galaxy
(e.g., \citealt{Stadel:2009el}). Gravity solvers which use hybrid
tree-particle-mesh techniques (e.g., {\sc{Gadget-2}},
\citealt{Springel:2005cza}) are ideally suited to carrying out such
calculations on these scales. In addition to tailored codes for
studying Milky Way sized halos, halo finders used for identifying
substructure contained within them have also drastically improved over
the past 30 years. Simple friends-of-friends (FoF) algorithms
(e.g., \citealt{Davis85}) have now evolved into parallel, fully
hierarchical FoFs algorithms adopting six phase-space dimensions and
one time dimension allowing shape-independent, and noise-reduced
identification of substructure (\citealt{Behroozi:2013cn}). These tools are very robust
methods for accurately identifying bound substructures
(e.g., \citealt{Onions:2012iv}), though \cite{2015arXiv150601405B} has recently highlighted the difficulty in connecting halos during merger events. These efforts demonstrate that only algorithms that combine phase-space \textit{and} temporal information should be
used.

Two primary groups have
performed zoom-in \textit{N}-body simulations of the growth of Milky Way sized
halos in extremely high resolution -- the \textit{Aquarius}
project of \cite{Springel:2008gd} and the \textit{Via Lactea}
simulations of \cite{Diemand:2008hr}. Whilst these works have been thoroughly successful and made it possible to
quantify the formation of the stellar halo, for example, both the \textit{Aquarius} and \textit{Via Lactea}
projects are limited in a number of respects.

The first of these is that they adopted the now observationally
disfavored \textit{Wilkinson Microwave Anisotropy Probe's} first set of
cosmological parameters (\textit{WMAP-1},
\citealt{Spergel:2003ci}). The advent of the \textit{Planck satellite}
(Planck \citealt{Collaboration:2014dt}) with three times higher resolution
and better treatment of the astrophysical foreground (owing in large
part to using nine frequency bands instead of five with
\textit{WMAP}) has allowed even more precise estimates of key
cosmological parameters. In particular, the most crucial of these for
accurate cosmological simulations are the baryon density ($\Omega_b$),
the matter density ($\Omega_c$), the dark energy density
($\Omega_\Lambda$), the density fluctuations at 8 \hMpc ($\sigma_8$)
and the scalar spectral index ($n_s$). \cite{Dooley:2014db} showed through a systematic studies of structure
formation using different cosmologies that the
maximum circular velocities, formation and accretion times of a given
host's substructure are noticeably different between cosmologies. $\sigma_8$ in \textit{WMAP-1} for example is
much higher ($\sigma_{8,WMAP1}$ = 0.9 vs. $\sigma_{8,Planck}$ = 0.83)
which shifts the peak in cosmic star formation rate to lower redshift,
resulting in slightly bluer galaxies at \textit{z} = 0
(\citealt{Jarosik:2011dy}, \citealt{Guo:2013jm},
\citealt{Larson:2015gk}).

The second major drawback and perhaps more significant is that the
\textit{Aquarius} and \textit{Via Lactea} simulations were simply
limited in number. The \textit{Aquarius} project consists of six
well-resolved Milky Way mass halos, while the \textit{Via Lactea}
study focused on only one such halo.

There exists significant halo-to-halo scatter in, e.g., the substructure shape and abundance owing to variations in accretion
history and environment, (\citealt{Springel:2008gd},
\citealt{Cooper:2010ita}, \citealt{BoylanKolchin:2010bd}), with the
dispersion appearing significant (a factor $> 3$,
\citealt{Lunnan:2012kt}). But based on such a small sample, the extent
cannot be well-quantified, although determining the distributions of
substructure properties of galaxy halos is critical for interpreting
the various observations of dwarf galaxy populations of all large
galaxies, including the Milky Way and Andromeda.

More recently, \cite{GarrisonKimmel:2014bca} have produced a suite of
36 Milky Way halos (24 isolated analogues, 12 Local Group analogues;
\textit{ELVIS} suite) at a resolution of $\sim$$10^5$ $M_\sun$ per
particle ($\sim$Aquarius level-3). Studies using this suite have
again highlighted the case for the \textit{too big to fail problem}
(\citealt{BoylanKolchin:2011tn}) by showing that
the so called ``massive failures'' (i.e. halos with \Vmax $\ge$ 25
\kms that became massive enough to have formed stars in the presence
of an ionizing background, \Vpeak $\ge$ 30 \kms) do not disappear when
larger numbers of halos across a range of host masses are simulated
\citep{GarrisonKimmel:2014bca}. Despite the \textit{ELVIS} suite's utility, it unfortunately
lacks the extra mass resolution required to study the formation of
minihalos and very small dwarf galaxies ($\sim10^6$\,M$_{\odot}$), both at the
present day and their evolution since the epoch of reionization. Also,
\textit{ELVIS} is not suitable\footnote{Particle tagging usually requires 1--5$\%$ of the most bound particles of a satellite to be tagged. For a simulation which resolves 10$^8$ \Msol hosts with $\sim$1000 particles (i.e.,\ $ELVIS$), this means one can only use  a single particle to contain all the baryonic information which is insufficient for modelling multiple stellar populations.} for using the \textit{particle tagging}
technique whereby a few per cent of the central dark matter particles of accreting systems are assigned
stellar properties to study the assembly of the stellar halo (e.g.,
\citealt{2015arXiv150104630C}). If we are to understand the origin
of the first stellar systems (including their chemical constituents) and to locate their descendants at the
present day, higher resolution as well as particle tagging is of
critical importance.

Whilst previous simulations all have their own merits and drawbacks,
one issue prevalent across nearly all previous studies is that they
introduced bias in selecting their halos. Usually halo candidates
studied using the zoom-in technique meet three criteria: isolation,
merger history and Lagrangian volume. From a computational standpoint,
if one can obtain a compact Lagrangian region, a quiet merger history
and keep the halo relatively isolated, the savings in CPU-hours can be
immense. Ultimately, however this three-pronged approach introduces a
selection bias. Whilst constructing a simulation with these three key
criteria in place will generate an approximate Milky Way analogue, one
will not gain an understanding of how the results from studying this
halo will compare to halos more generally selected from a pool in the
desired mass range (e.g., 1\ -- 2 $\times$ $10^{12}$ \Msol).

The first requirement is that the halos have a quiescent merger
history, which is usually defined by the host having no major merger
since a given redshift, e.g., \textit{z} = 1
(e.g., \citealt{Springel:2008gd}). Constraining the merger history of a simulation suite severely limits the capabilities of reconstructing the formation history of the Milky Way.  Indeed, by statistically contrasting observational data sets to mock data extracted from a set of Milky Way-like dark matter halos,  coupled to a semi-analytical model of Galaxy formation \citep{Tumlinson:2009ia}, \cite{Gomez:2012hh} showed the best-fitting input parameter selection strongly depends on the underlying merger history of the Milky Way-like galaxy. For example, even though for every dark matter halo it is always possible to find a best-fitting model  that  tightly reproduces the Milky Way satellite luminosity function, these best-fitting models generally fail to reproduce a second and independent set of observables (see \citealt{Gomez:2014fp}). It is thus critical to sample a wide range of evolutionary histories.  The second requirement that the
Lagrangian volume of the halo's particles be compact also in part
biases the merger history of the halo. For a fixed \textit{z} = 0
virial mass, the smaller the Lagrangian volume of a halo, the less
likely that halo will have a late major-merger event. This bias
further compounds the aforementioned issues of selecting halos with
quiet merger histories. Lastly, the isolation criteria preferentially
selects halos in low density environments, resulting in decreased
substructure \citep{RagoneFigueroa:2007ei} and higher angular momentum
(\citealt{AvilaReese:2005io}, \citealt{Lee:2006hq}).

In light of all of these issues, we are motivated to create a
comprehensive dataset consisting of 60--70 dark matter halos of approximately Milky Way mass in extremely
high spatial and temporal resolution with a more relaxed selection
criteria to not just understand the origin and evolution of the Milky
Way, but additionally how it differs to other galaxies of similar mass
\textit{in general}. Moreover, this new simulation set (unlike the $Aquarius$
and $Via$ $Lactea$ which were very specific in nature) lends itself well to studying the substructure and
stellar halos of $\sim$10$^{12}$ \Msol galaxies such as those being
studied in the recently completed {\sc{ghosts}} survey \citep{deJong:2007kq,Monachesi:2013fz,2015arXiv150706657M}. 

We call this simulation suite \textit{The Caterpillar Project} owing to
the similarity between each of the individual halos and how they work
together towards a common purpose. Due to the extreme computational
requirement for a project of this size ($\sim$14M CPU hours and
$\sim$700TB of storage), we are staggering our release. For this first
paper, we focus on the general $z$ = 0 properties of the first 24
halos of the \textit{Caterpillar} suite in order to clearly
demonstrate data integrity and utility. In Section 2, we outline the simulation suite parameters, numerical
techniques, and halo properties. In Section 3 we present a variety of
initial results drawn from the suite. In Section 4, we present our primary conclusions
from our initial subset of halos. Lastly, we present an Appendix with
details of our convergence study and parameters used in the construction of our initial conditions.

\section{The Caterpillar Suite}
\label{sec:methods}

\subsection{Simulation $\&$ Numerical Techniques}

The \textit{Caterpillar} suite was run using {\sc{P-Gadget3}} and {\sc{Gadget4}}, tree-based $N$-body codes based on {\sc{Gadget2}} (\citealt{Springel:2005cza}). For the underlying cosmological model we adopt the $\Lambda$CDM parameter set characterised by a \textit{Planck} cosmology given by, $\Omega_m=0.32$, $\Omega_\Lambda=0.68$, $\Omega_b=0.05$, $n_s=0.96$, $\sigma_8=0.83$ and Hubble constant, H = 100 $h$ km s$^{-1}$ Mpc$^{-1}$ = 67.11 km s$^{-1}$ Mpc$^{-1}$ (\citealt{Collaboration:2014dt}). All initial conditions were constructed using {\sc{music}} (\citealt{Hahn:2011gj}).  We identify dark matter halos via \rockstar (\citealt{Behroozi:2013cn}) and construct merger trees using {\sc{Consistent-Trees}} (\citealt{Behroozi:2012dz}). \rockstar assigns virial masses to halos, \Mvir, using the evolution of the virial relation from \cite{Bryan:1998cc} for our particular cosmology. At \textit{z} = 0, this definition corresponds to an over-density of 104 $\times$ the critical density of the Universe. We have modified \rockstar to output \textit{all} particles belonging to each halo so we can reconstruct any halo property in post-processing if required. We have also improved the code to include iterative unbinding (see Section \ref{sec:updaterockstar}). In this work, we restrict our definition of virial mass to include only those particles which are bound to the halo.

\subsection{Parent Simulation, Zoom-ins $\&$ Contamination}
\label{sec:halos}

\begin{figure}[!h]
\includegraphics[width=0.48\textwidth]{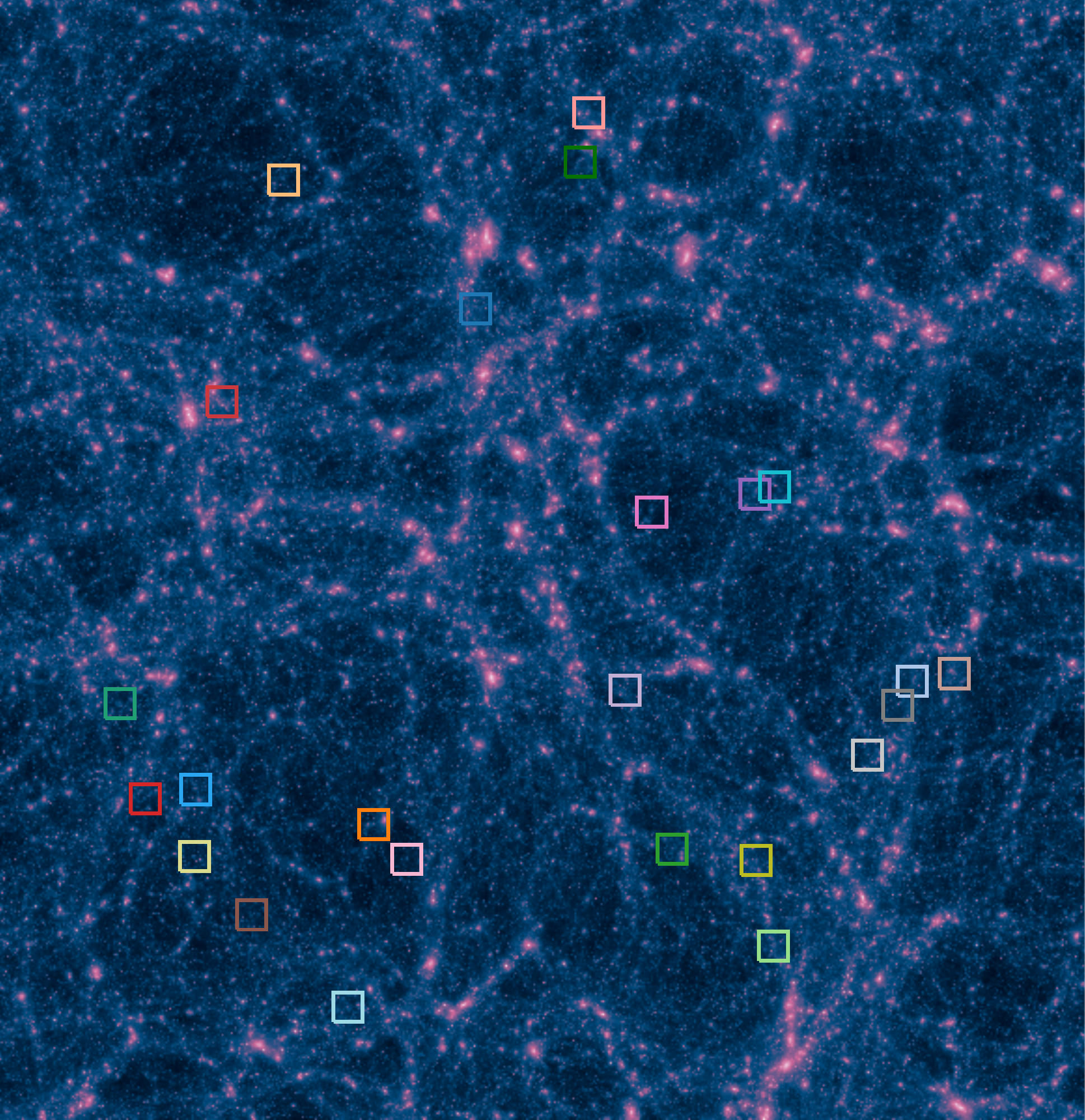}
\caption{Projected dark matter density at  $z$ = 0 of the parent simulation, from which the 70 Caterpillar halos were selected. The box width is 100 \hMpc and the color represents the logarithm of the dark matter density. The colored circles correspond to the location of the first 24 $Caterpillar$ halos. The color for a given halo is kept identical for all figures throughout this work.}
\label{fig:parentsim}
\end{figure}

Initially a parent simulation box (see Fig.\ \ref{fig:parentsim}) of width 100 $h^{-1}$ Mpc was run at $N_{p}\ =\ 1024^3$ ($m_{p} = 8.72 \times 10^7\ h^{-1}$ M$_\sun$) effective resolution (see {\sc{music}}/{\sc{P-Gadget3}} parameter files on project website) to select viable candidate halos for re-simulation (i.e.,\ $\sim$10,000 particles per host). The candidates for re-simulation were selected via the following mass and isolation criteria:

\begin{itemize}
\item Halos were selected between 0.7 $\times$ 10$^{12}$ M$_\sun$ $\leq$ \Mvir $\leq$ 3 $\times$ 10$^{12}$ M$_\sun$ (\citealt{Smith:2007fi}, \citealt{Xue:2008kb}, \citealt{Tollerud:2012hr}, \citealt{BoylanKolchin:2013et}, \citealt{Sohn:2013do}, \citealt{Piffl:2014hd}, \citealt{2015arXiv150308508G}, \citealt{2015arXiv150703594P}, see \citealt{2015MNRAS.453..377W} for review).
\item No halos with \Mvir $\ge$ 7 $\times \ 10^{13}$ M$_\sun$ within 7 Mpc (\citealt{Li:2008ji}. \citealt{vanderMarel:2012jaa}). We avoid halos near large clusters which would greatly enhance our Lagrangian volumes, making our ability to run simulations at our desired resolution impossible.
\item No halos with \Mvir $\ge$ 0.5 $\times$ $M_{host}$ within 2.8 Mpc (\citealt{Karachentsev:2004dm}, \citealt{Tikhonov:2009eh}). We currently avoid pairs in our sample owing to the difficulty of running them at our desired resolution at the present time. We have nevertheless selected equivalent pairs of our current isolated sample but will be examining those in future work.
\end{itemize}

This results in 2122 candidates being found (from an original sample of 6564 within the specified mass range). We use an extremely weak selection over merger history such that we require no halo to have had a major merger (1:3 mass ratio) since $z$ = 0.05 ($<$5$\%$). Our overall aim is to construct a representative sample of 10$^{12}$ \Msol halos and not specifically require Milky Way analogues \textit{a priori} as has been done in previous studies (\citealt{Diemand:2007fi}, \citealt{Stadel:2009el}, \citealt{Springel:2008gd}, \citealt{GarrisonKimmel:2014bca}, \citealt{2014arXiv1412.2748S}). This also allows us to apply statistical tools to constrain semi-analytic models in future work (e.g.,\ \citealt{Gomez:2014fp}). We place our halos into three mass bins with the largest number of halos centered on the most likely mass for the Milky Way ($M_{200}$ = $1.6^{+0.5}_{-0.4} \times 10^{12}$ \Msol, \citealt{Piffl:2014hd});

\[ M_i = \left\{ \begin{array}{ll}
                0.7 \-- 1.0 \times 10^{12}\ \mathrm{M_\sun\xspace}: 7 \ \mathrm{halos}\\
                1.0 \-- 2.0  \times 10^{12}\ \mathrm{M_\sun\xspace}: 46 \ \mathrm{halos}\\
                2.0 \-- 3.0  \times 10^{12}\ \mathrm{M_\sun\xspace}: 7 \ \mathrm{halos}
        \end{array} \right. \] 

For this paper we are only considering a subset of the total sample in preparation, specifically 21 halos within the 1 -- 2  $\times$ 10$^{12}$ \Msol mass range and 3 halos within the $0.7 \-- 1.0 \times 10^{12}$ \Msol mass range.

\subsection{Contamination Study}
As has been highlighted by \cite{Onorbe:2014kp}, a great deal of care has to be taken when carrying out re-simulations of this kind so as to avoid contamination of the main halo of interest by low-resolution particles at $z$ = 0. If mass from low resolution particles contributes more than $\sim$$2\%$ of the total host mass there can be offsets to estimates of the halo profile, shape, spin, and especially gas properties in hydrodynamic runs. To avoid contamination in our sample we custom built a Python GUI (using {\sc{TraitsUI}}), \href{http://github.com/bgriffen/cme}{\textit{Caterpillar Made Easy}} ({\sc{cme}}), for running and analyzing cosmological simulations (for both single and multi-mass simulations). This tool allowed us to carry out an extensive contamination study (i.e.,\ using $\sim$264 low resolution test halos with a particle mass of $\sim10^7$ \Msol) specifically for the halos to be re-simulated. We have $automated$ the monotony of constructing hundreds of qualitatively similar $but$ quantitatively distinct cosmological simulations with the added benefit of being able to interactively select over initial condition parameters, cosmologies, halo finders and merger trees. This procedure was carried out self-consistently across all runs allowing for a systematic study of which simulation parameters produce the most computationally inexpensive to run, uncontaminated halos.

Using {\sc{cme}} we tested eleven Lagrangian geometries (e.g.,\ convex hull, ellipsoid, expanded ellipsoids, cuboids and expanded cuboids) so as to ensure a sphere of radius $\sim$1 \hMpc exists of purely uncontaminated (high-resolution) particles centered on the host halo at $z=0$. Our need to run eleven different Lagrangian geometries for each halo is motivated by the fact that the geometries vary substantially from halo to halo (due partially to their varied merger histories) and we wished to minimize the computation cost whilst achieving our contamination goals. It must also be highlighted that unlike many other studies, we did not select one Lagrangian geometry for all halos but used a specific geometry for a given halo depending on the needs of its simulation.

\begin{figure}
\includegraphics[width=0.23\textwidth]{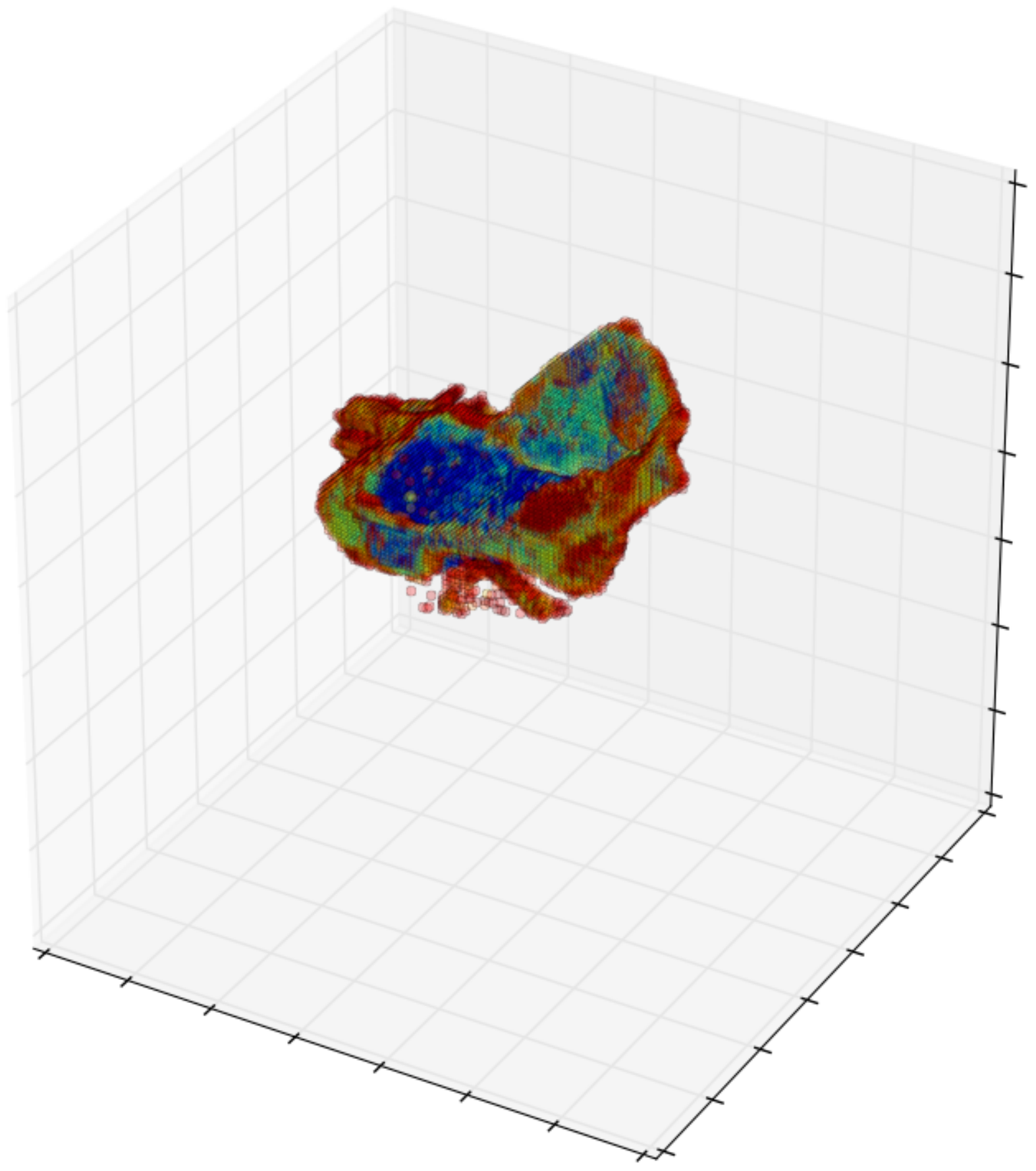}
\includegraphics[width=0.23\textwidth]{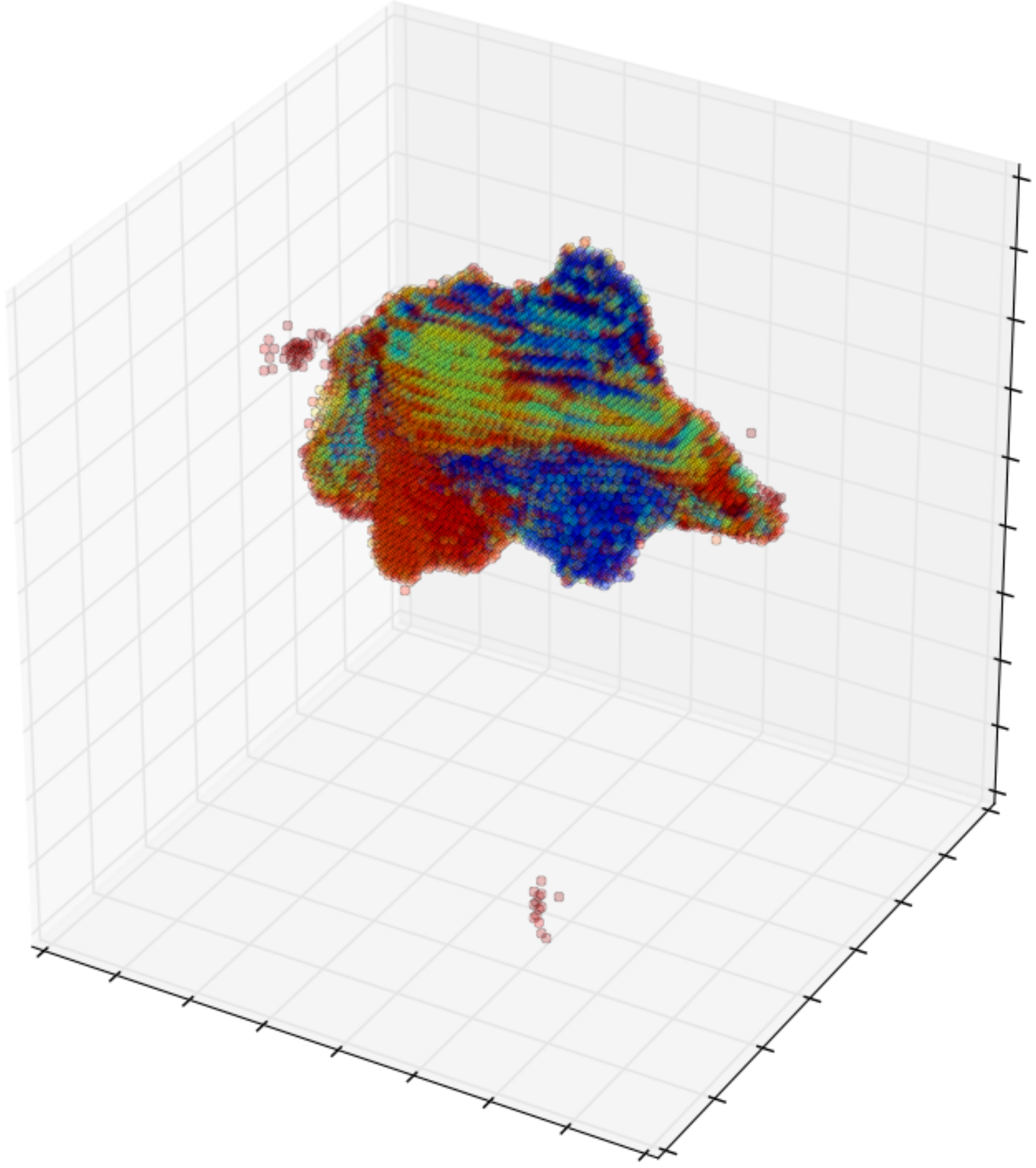}
\includegraphics[width=0.23\textwidth]{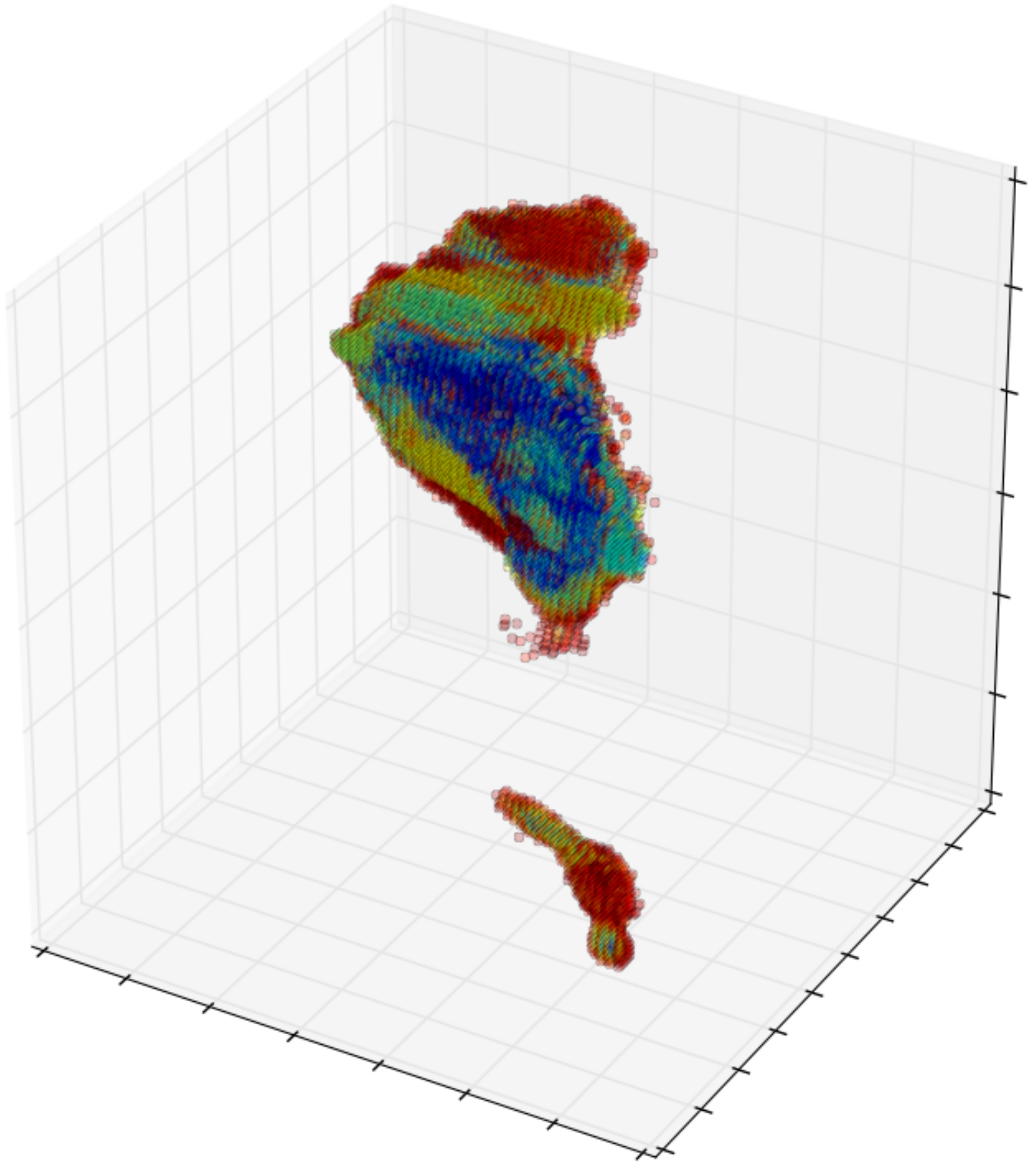}
\includegraphics[width=0.23\textwidth]{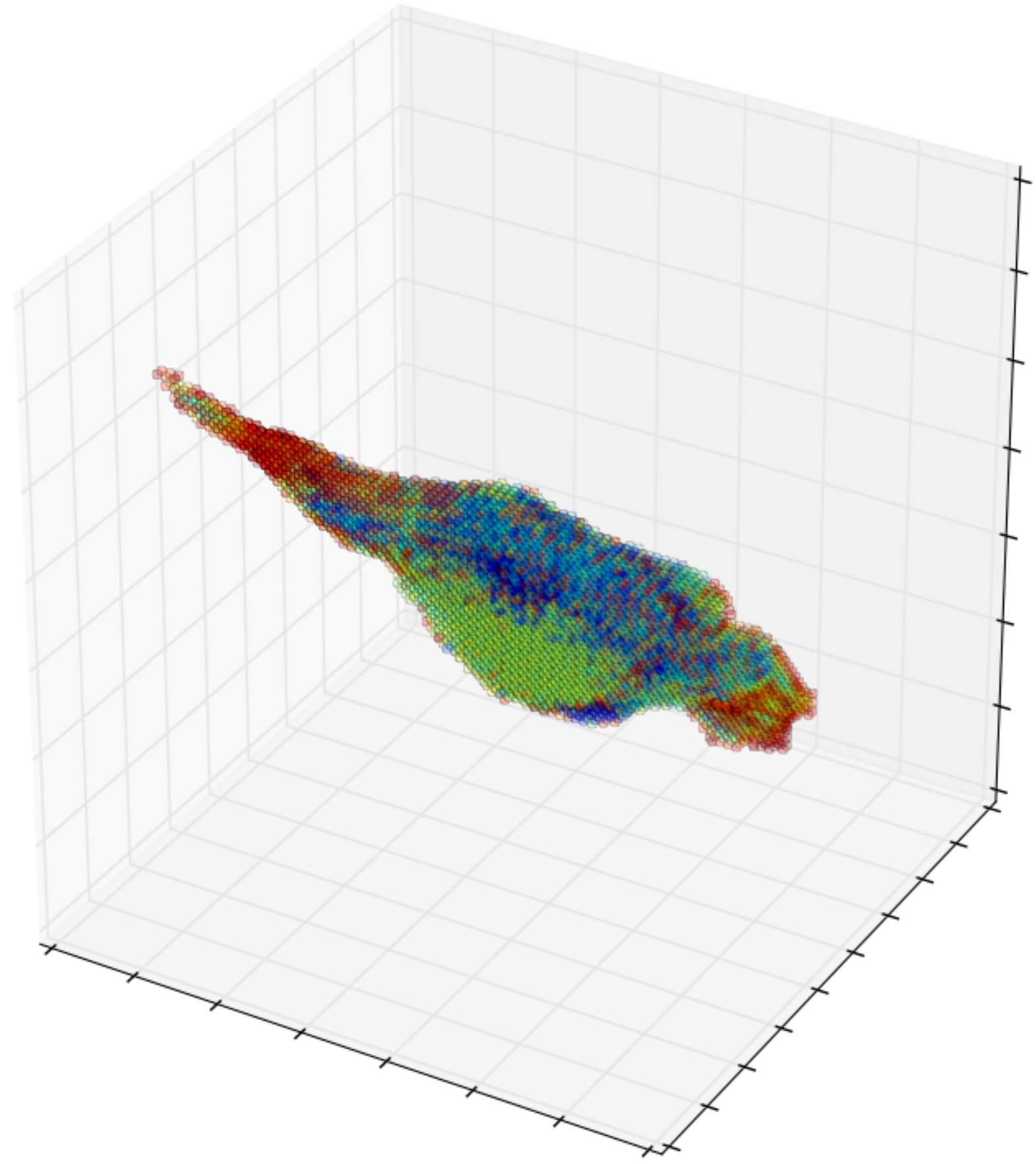}
\caption{Sample Lagrangian volumes (at $z$ = 127) of halos from the parent simulation. Some geometries are easily bounded by a minimum cuboid (e.g.,\ top left) but others require an ellipsoid (e.g.,\ bottom right). In some cases, there reside particles well away from the primary Lagrangian volume (e.g.,\ top right and bottom left). For these difficult situations, we carefully trimmed the Lagrangian geometry to ensure no contamination of low resolution particles within 1 \hMpc of the host at  $z$ = 0. Particles are color coded by distance from the host at  $z$ = 0 where red particles are 5$\times$\Rvir and blue particles are within the virial radius.}
\label{fig:lagrvol}
\end{figure}

In Table \ref{tab:icgeom} we show the various geometries we used for constructing our initial conditions. We modified {\sc{music}} to be able to produce expanded Lagrangian volumes rather than the bounded volumes with which it was originally published. In Figure \ref{fig:lagrvol} we show four examples of Lagrangian geometries for halos selected for re-simulation. In some cases the geometries are reasonably compact allowing for the traditional minimum cuboid enclosing to be used. Some larger regions however are extremely non-spherical (e.g.,\ bottom right  of Fig. \ref{fig:lagrvol}) and so a minimum ellipsoid was used. For each geometry we take the enclosed volume at  $z$ = 0 denoted by either $4 \times R_{{\rm vir}}(z=0)$ and $5 \times R_{{\rm vir}}(z=0)$. We run each of these halos to $z$ = 0, run our modified \rockstar and determine at what distance the closest low resolution or contamination particle (type = 2) resides in each case. With the knowledge that the high-resolution volume distance decreases at higher levels of refinement \citep{Onorbe:2014kp}, we ensure a minimum contamination distance of $\sim$1 \hMpc at $z$ = 0 at our lowest resolution re-simulation with the desire to have uncontaminated spheres of radius, $\sim$1 Mpc at our highest resolution re-simulation. In cases where four times the virial radius enclosure created contaminated halos but five times the virial radius created too large a simulation to run, we opted for an expanded ellipsoid of the minimum enclosing ellipsoid. In some cases there were a handful of offending particles far away from the primary Lagrangian volume (e.g.,\  Figure \ref{fig:lagrvol} top right and bottom left) making no standard geometry enclosure feasible, expanded or otherwise. Here we trimmed the Lagrangian volume by hand and simulated the new geometry to $z$ = 0 to ensure it had no contamination. Traditionally these types of halos are avoided but since we do not want to bias our sample, we dealt with complicated geometries in this specialized manner and have included them in our sample. Using this tailored approach, our highest resolution runs obtain very large, high-resolution regions with spheres of radius $\sim$$1.4 \pm 0.4$ Mpc of solely high-resolution particles.

\begin{figure}
\includegraphics[width=0.48\textwidth]{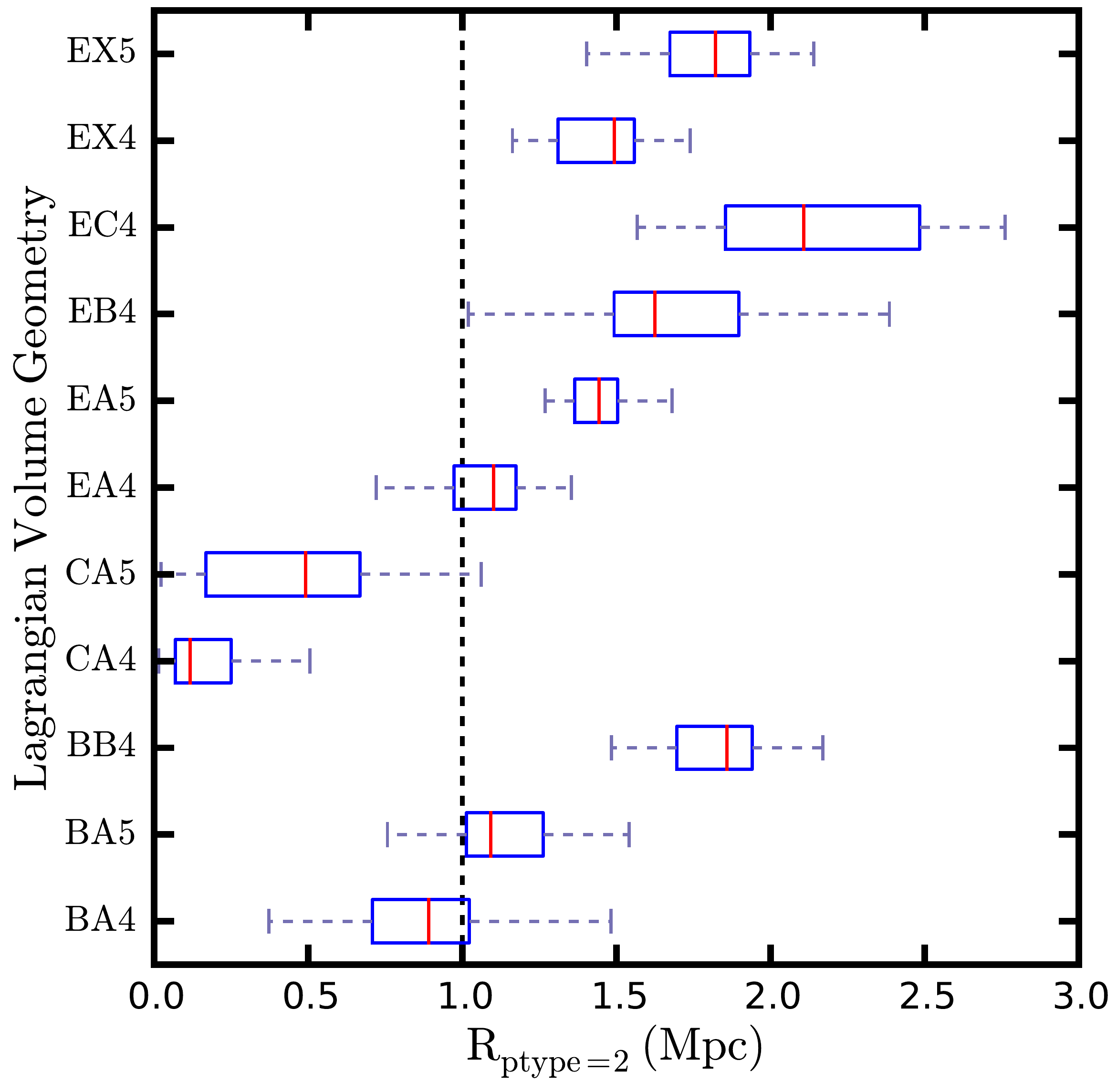}
\caption{Box plot of the distances to the first low resolution particle (particle type 2) for each of the prospective initial condition geometries (i.e.,\ all run at {\sc{levelmax}} = 11 in {\sc{music}}, one level above our parent simulation). Table \ref{tab:icgeom} contains the key for each geometry. The red lines indicate the median, edges of the box represent 25$\%$ and 75$\%$ quartiles and the outer tick marks represent the maximum and minimum distance. The dashed line represents the threshold we used to determine if a geometry was viable for a higher level re-simulation, though this was balanced against computational cost. We select the geometry which used the fewest CPU hours but maintained the largest uncontaminated volume.}

\label{fig:contamr}
\end{figure}

In Figure \ref{fig:contamr} we show box plots of the median contamination distance and respective quartiles for all 264 of our test halos using each of our selected geometries. Typically, the best performing geometry (i.e.,\ the largest uncontaminated volume with the cheapest computational expense) was the expanded ellipsoid which enclosed all particles within 4 or 5 times the virial radius of the host in the parent simulation at $z = 0$.

\begin{figure*}
\includegraphics[width=0.24\textwidth]{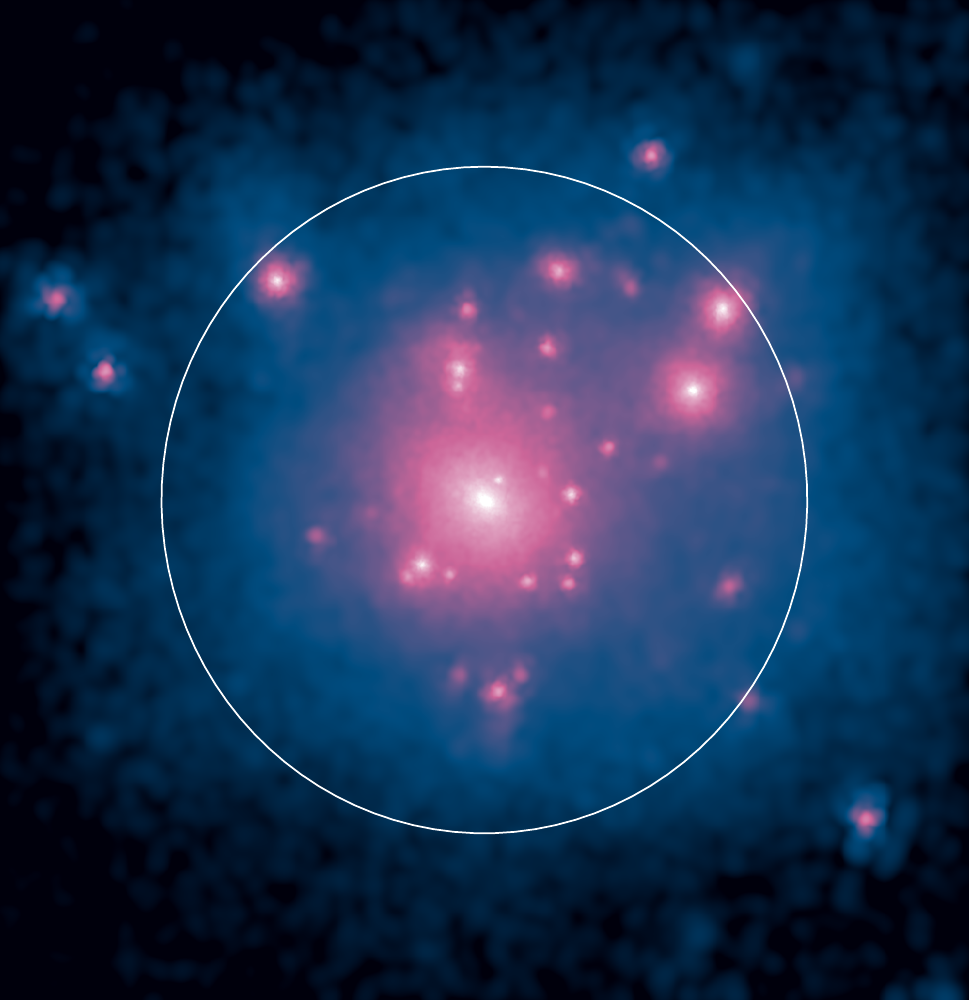}
\includegraphics[width=0.24\textwidth]{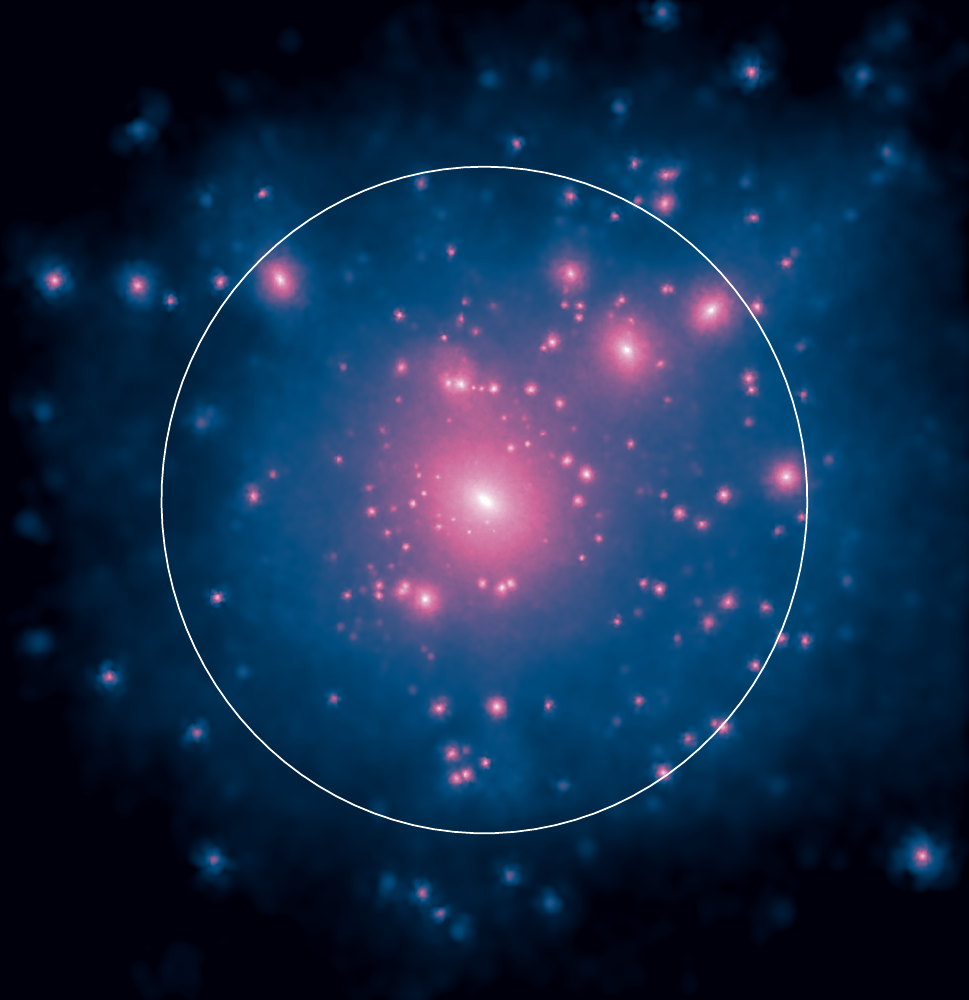} 
\includegraphics[width=0.24\textwidth]{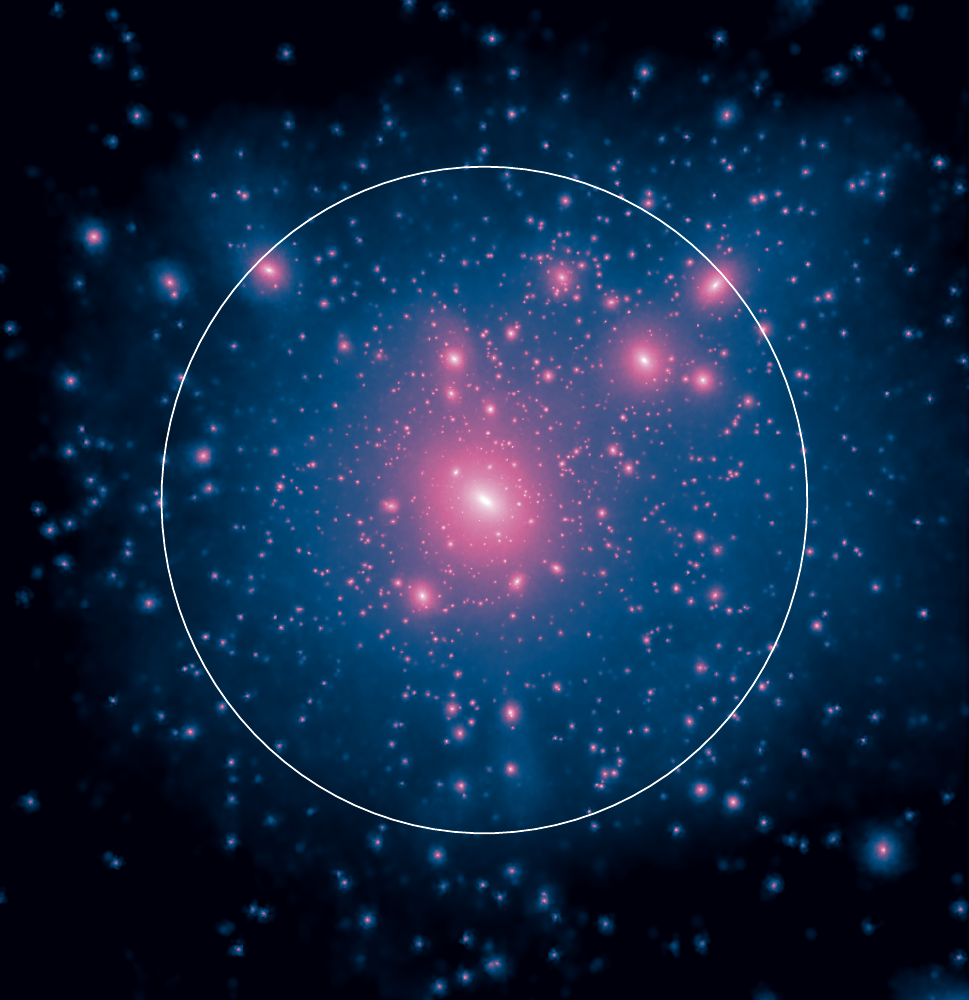} 
\includegraphics[width=0.24\textwidth]{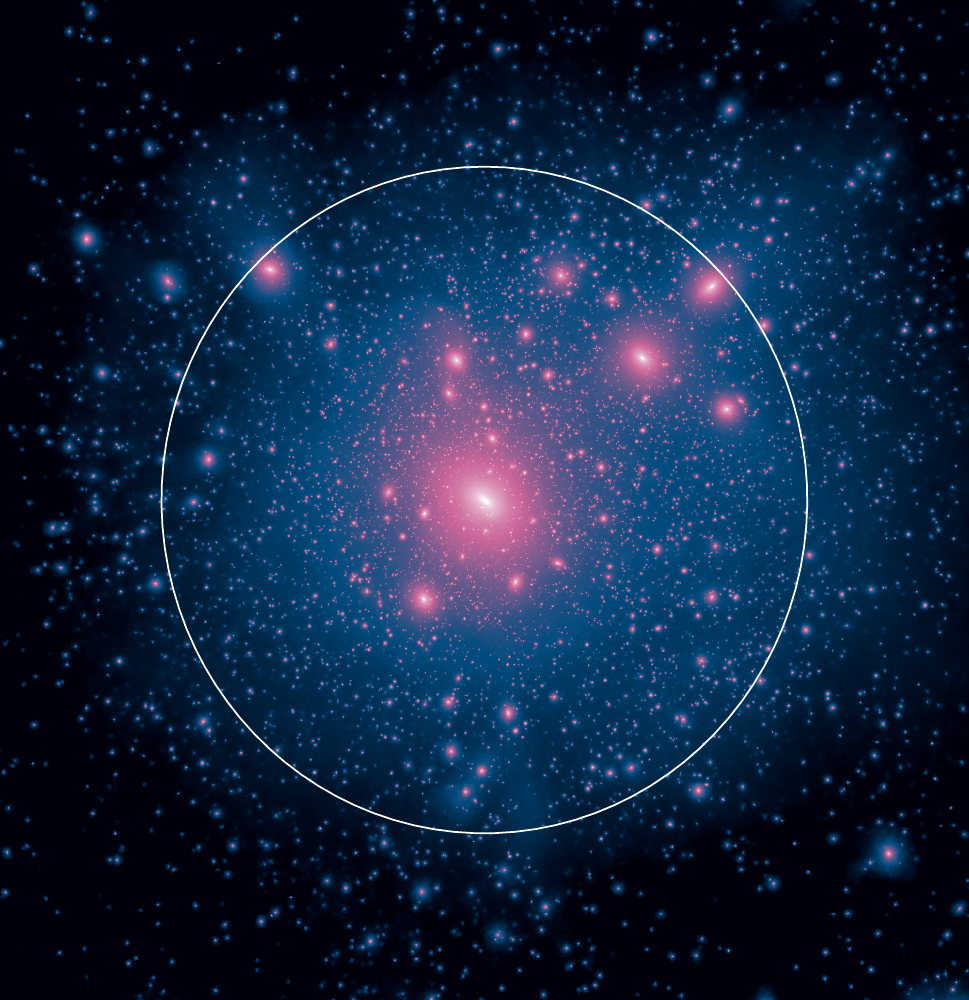} 
\caption{Projected dark matter density at  $z$ = 0 of Cat-1 at successively higher resolutions (increasing by a factor of 8 in mass resolution each time) from left to right. The left panel is {\sc{lx}} = 11 ($m_p = 1.53\times10^8 M_\odot$) and  the right most panel is {\sc{lx}} = 14 ($m_p = 2.99\times10^4 M_\odot$). The white circle represents the virial radius, \Rvir. The image brightness is proportional to the logarithm of the dark matter density squared (i.e.,\ log($\rho^2$).}
\label{fig:resolutioncheck}
\end{figure*}

\begin{deluxetable}{cccc}
\tablecolumns{4}
\tablecaption{The contamination suite used on the first refinement level (i.e.,\ {\sc{levelmax}} =\ 11) for every halo in the $Caterpillar$ suite.}
\tablehead{  
 \colhead{Name}  & \colhead{$nR_{vir}(z=0)^a$} & \colhead{Geometry}   &  \colhead{Factor$^b$}}
\startdata
CA4 & 4 & Convex Hull & --\\
CA5 & 5 & Convex Hull  & --\\
EA4 & 4 & Bounded Ellipsoid & -- \\
EA5 & 5 & Bounded Ellipsoid & -- \\
EX4 & 4 & Expanded Ellipsoid & 1.05 \\
EX5 & 5 & Expanded Ellipsoid & 1.05  \\
EB4 &  4 & Expanded Ellipsoid & 1.1  \\
EC4 &  4 & Expanded Ellipsoid & 1.2  \\
BA4 & 4 & Minimum Cuboid & --  \\
BA5 & 5 & Minimum Cuboid& --  \\
BB4 & 4 & Expanded Cuboid & 1.1  
\enddata
\label{tab:icgeom}
\tablenotetext{a}{The multiple of the $z=0$ virial radius that we used to construct the Lagrangian volume.}

\tablenotetext{b}{The factor we increased the original volume (e.g.,\ 1.2 means the ellipsoid was expanded by 20$\%$ in size. A dash represents the minimum ellipsoid/cuboid/hull exactly). These values were arbitrarily chosen with the only requirement being that the initial condition files were not overly large in size (i.e.,\ a few hundred megabytes at {\sc{lx11}}).}
\end{deluxetable}

\newpage
\subsection{Zoom-in Simulations}
Starting from our parent simulation resolution, we re-simulated each halo at iteratively higher resolutions (a factor of 8x increase in particle number for each level) to ensure we did not obtain contaminated particles within the host halo (the uncontaminated volume shrinks with an increase in the ratio of the zoom-in resolution to that of the parent simulation resolution). In Figure \ref{fig:resolutioncheck}, we show the dark matter distributions for iteratively higher resolutions of the same halo. One can clearly identify the same subhalos across all resolutions indicating the qualitative success of our numerical techniques. Regarding computational resources, each halo at our highest resolution took between 150 -- 300K hours on TACC/Stampede  and occupy $\sim$5 -- 10\ TB of storage for both the raw {\sc{hdf5}} snapshots and halo catalogue. Table \ref{tab:sim} shows the mass and spatial resolution for each of our refinement levels. Our softening length is $\epsilon \sim$ 76 \hpc for our fiducial resolution.

\begin{deluxetable}{cccc} 
\tablecolumns{4}
\tablecaption{The resolution levels of the \textit{Caterpillar} suite. \label{tab:sim}}
\tablehead{ 
   \colhead{{\sc{lx}}} & 
  \colhead{$N_p$} & 
  \colhead{$m_p$ } & 
  \colhead{$\epsilon$ } \\
  & & \colhead{(\Msol)} & \colhead{(\hpc)}
}
\startdata

{\sc{15}} & 32768$^3$ & $3.7317 \times 10^3$ & 38 \\
{\sc{14}} & 16384$^3$ & $2.9854 \times 10^4$ & 76 \\
{\sc{13}} & 8096$^3$ &   $2.3883 \times 10^5$ & 152    \\
{\sc{12}} & 4096$^3$ &   $1.9106 \times 10^6$ & 228  \\
{\sc{11}} & 2048$^3$ &   $1.5285 \times 10^7$ & 452 \\
\hline
{\sc{10}} & 1024$^3$ &   $1.2228 \times 10^8$ & 904 
\enddata

\tablecomments{{\sc{lx}} represents represents the effective resolution ($N_p = (2^X)^3$) of the high resolution region given by parameter, {\sc{levelmax}} in {\sc{music}}. $m_p$ is the particle mass and $\epsilon$ is the Plummer equivalent gravitational softening length. The parent simulation parameters are also shown in the last row. Only a select sample of halos are run at resolution level {\sc{lx15}}. These runs will be presented in future works.}
\end{deluxetable}

We space our snapshots (320 per simulation) in the logarithm of the expansion factor until $z$ = 6 ($\sim$5 Myrs/snapshot) and then linear in expansion factor down to $z$ = 0 ($\sim$50 Myrs/snapshot). The motivation for this piecewise stitching of the two temporal schemes is two-fold. At $z$ $>$ 6, we enter the era of mini-halo formation and the reionization epoch. If we wish to model the transport of Lyman-Werner (LW) radiation semi-analytically from the first mini-halos (e.g.,\ \citealt{Agarwal:2012ik}), we require a temporal resolution on par with the mean free path of LW photons in the intergalactic medium and the lifetime of a massive Population III star ($\leq$ 10 Myrs). Secondly, we also wish to resolve the disruption of low mass dwarf galaxies at low redshift, which requires a temporal resolution of order $\sim$ 50 Myrs (e.g.,\ Segue I has a disruption time scale of $\sim$ 50 Myrs). These time scales are also required if one is attempting to determine subhalo orbital pericenters which can be input into semi-analytic models of tidal disruption (e.g.,\ \citealt{Baumgardt:2003dl}). While we intend to use the capabilities offered by finely sampled snapshots in future work, this initial paper primarily focuses on the  $z$ = 0 halo properties.

\subsection{Iterative Unbinding In \rockstar}
\label{sec:updaterockstar}
\rockstar is able to find any overdensity in 6D phase space including both halos and streams. To distinguish gravitationally bound halos from other phase space structures, \rockstar performs a single-pass energy calculation to determine which particles are gravitationally bound to the halo. Over-densities where at least 50$\%$ of
the mass is gravitationally bound are considered halos, with the exact fraction a tuneable parameter (\unboundthreshold) of the algorithm \citep{Behroozi:2013cn}.

This definition is generally very effective at identifying halos and subhalos -- but it fails in two important situations.
First, if a subhalo is experiencing significant tidal stripping, the 50$\%$ cutoff can remove a subhalo from the catalog that should actually exist. We have found that changing the cutoff can recover the missing subhalos, but the best value of the cutoff is not easily determined.
Second, \rockstar is occasionally \textit{too} effective at finding substructure in our high resolution simulations. In particular, it often finds velocity substructures in the cores of our halos that are clearly spurious based on their mass accretion histories and density profiles.
Importantly, these two issues do not just affect low mass subhalos, but they can also add or remove halos with \Vmax $> 25$ \kms.

\begin{figure}[h]
\includegraphics[width=0.48\textwidth]{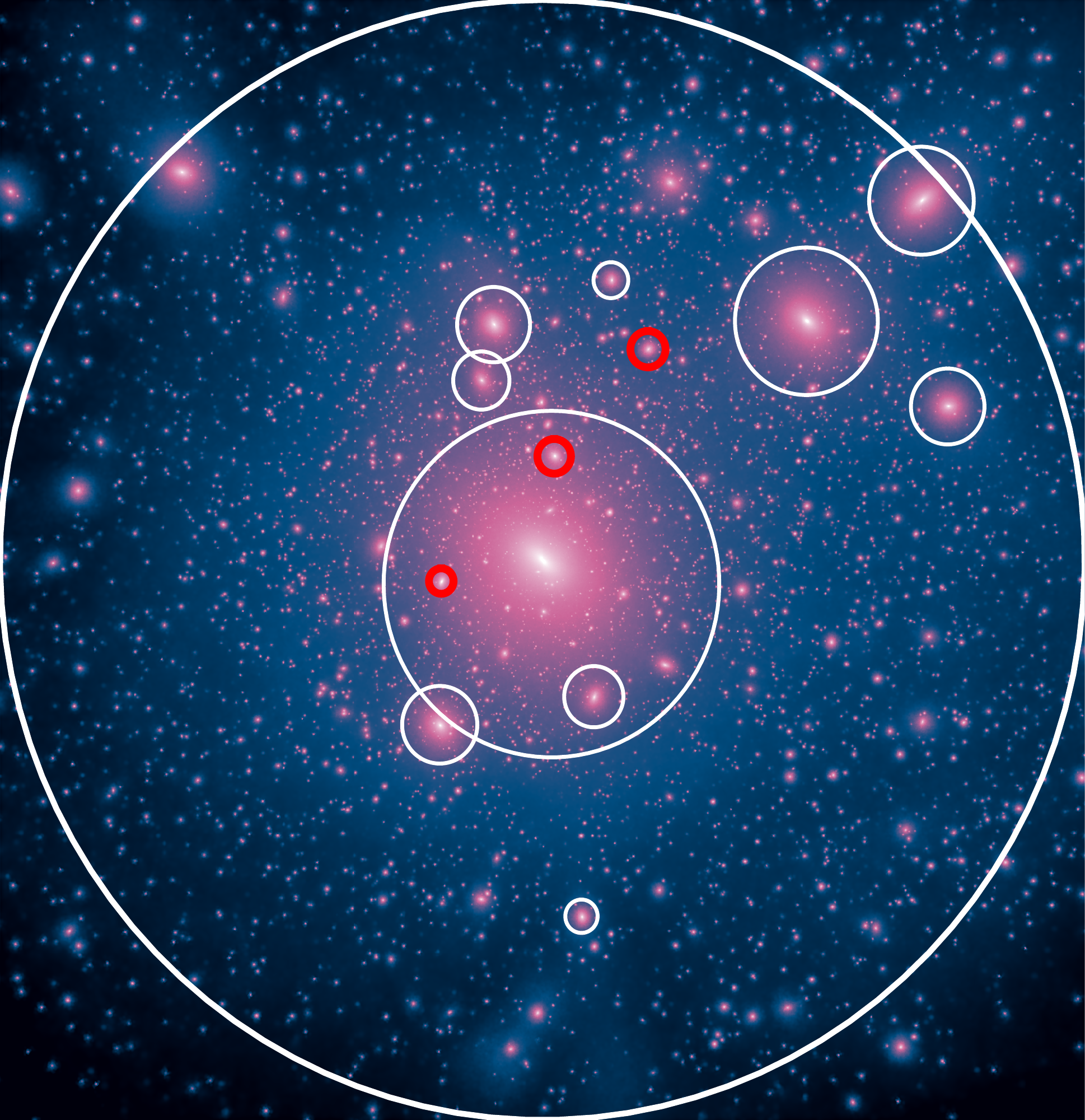} 
\caption{Density projection of the Cat-1 halo with subhalos with \Vmax $>$ 30 \kms highlighted by circles. The size of the circles corresponds to 3$\times$\Rmax (i.e.,\ the radius at which the velocity profile reaches its maximum). The white circles indicate subhalos found by the default parameters of the halo finding algorithm \rockstar (i.e.,\ without \textit{iterative} unbinding). The three red circles indicate halos which are recovered when complete iterative unbinding is used. The largest outer white circle indicates the virial radius of the host halo.}
\label{fig:iterunbind}
\end{figure}

Both of these problems can be alleviated by applying an iterative unbinding procedure. We have implemented such an iterative unbinding procedure within \rockstar. At each iteration, we remove particles whose kinetic energy exceeds the potential energy from other particles in that iteration. The potential is computed with the \rockstar Barnes-Hut method (see Appendix B of \citealt{Behroozi:2013cn}). We iterate the unbinding until we obtain a self-bound set of particles. Halos are only considered resolved if they contain at least 20 self-bound particles. All halo properties are then computed as usual, but with the self-bound particles instead of the one-pass bound particles. The iterative unbinding recovers the missing subhalos and removes most but not all of the spurious subhalos. Across 13 of our $Caterpillar$ halos, we recover 52 halos with subhalo masses above 10$^8$ \Msol which would have otherwise been lost using the conventional \rockstar. Figure \ref{fig:iterunbind} demonstrates how these large haloes can be recovered when our iterative unbinding procedure is used.

To remove the remaining spurious subhalos, we also remove halos if \Rmax (i.e.,\ the radius at which the velocity profile reaches its maximum) of the subhalo is larger than the distance between the subhalo and host halo centers. The downside to adding iterative unbinding is that it increases the run time for \rockstar by $\sim$50$\%$. In the rest of this paper, we only consider subhalos with at least 20 self-bound particles passing the \Rmax cut. We define the subhalo mass, \Msub, as the total gravitational bound mass of a subhalo which is obtained after the complete iterative unbinding procedure has been carried out.

\section{Results}
\label{sec:results}

\subsection{Host Halo Properties}
\label{sec:halos}

In Table \ref{tab:catsuite}, we provide the basic properties of our first 24 re-simulated halos. This includes the simulation name, the halo virial mass, the halo virial radius, concentration, maximum circular velocity, the radius at which the maximum circular velocity occurs, the formation time (defined as when the halo reaches half its present day mass), the redshift of the last major merger (1:3 mass ratio), the fraction of the host mass contained within subhalos, the axis ratios defining the halo shape, and the distance to the closest contamination particle from the host. We adopt a simple naming convention based on when the halos were post processed (1 -- ${\rm N_{halos}}$). Where required, we use a shorthand reference to the resolution of the simulation. These refer to the parameter {\sc{levelmax}} inside the IC generation code {\sc{music}} (e.g.,\ {\sc{levelmax}} = 14 is simply {\sc{lx14}}). This means {\sc{lx14}} represents an effective resolution of $N_{p} = (2^{14})^3$ ($\sim Aquarius$ level-2), {\sc{lx13}} $\rightarrow$ $N_{p} = (2^{13})^3$ ($\sim Aquarius$ level-3 or $\sim$$ELVIS$ resolution), {\sc{lx12}} $\rightarrow$ $N_{p} = (2^{12})^3$ and {\sc{lx11}} $\rightarrow$ $N_{p} = (2^{11})^3$. Unless otherwise stated, all halos in the analysis of this paper are the {\sc{lx14}} halos (i.e.,\ our flagship resolution). All halos have similar $z$ = 0 properties except Cat-7 whose properties can be tied to the fact that it has recently undergone a massive merger (1:3 mass ratio at $z$ = 0.03). We obtain extremely large uncontaminated volumes ($\sim$1.4 Mpc) in all but one of our halos (Cat-18 is $\sim$$3\%$ contaminated by mass). The fraction of mass held in subhalos across our sample is $f_{m,subs} = 0.121 \pm 0.041$ (1$\sigma$), though this excludes Cat-7.
\begin{deluxetable*}{cccccccccccc}
\tablecolumns{10}
\tablecaption{The halo properties of the first 24 $Caterpillar$ halos. \label{tab:catsuite}}
\tablehead{  
 \colhead{Halo} &  
 \colhead{\Mvir} & 
 \colhead{\Rvir} &  
 \colhead{c \tablenotemark{$a$}} &  
 \colhead{\Vmax} &  
 \colhead{$R_{\mathrm{max}}$ \tablenotemark{$b$}} & 
  \colhead{$z_{form}$ \tablenotemark{$c$}} &  
  \colhead{$z_{lmm}$ \tablenotemark{$d$}} & 
  \colhead{$f_{\mathrm{m,subs}}$ \tablenotemark{$e$}} &  
   \colhead{$c/a$} &  
   \colhead{$b/a$} &  
  \colhead{$R_{\mathrm{hires}}$ \tablenotemark{$f$}}  \\  
 \colhead{Name}  & \colhead{($\times 10^{12}$ \Msol)} & \colhead{(kpc)}   &  &  \colhead{(\kms)} &  \colhead{(kpc)}  & &  &  &  &  &    \colhead{(Mpc)} }
\startdata
Cat-1 & 1.559 & 306.378 & 7.492 & 169.756 & 34.083 & 0.894 & 2.157 & 0.207 & 0.841 & 0.869 & 0.998 \\
Cat-2 & 1.791 & 320.907 & 8.374 & 178.851 & 55.268 & 0.742 & 0.731 & 0.148 & 0.636 & 0.719 & 1.463 \\
Cat-3 & 1.354 & 292.300 & 10.170 & 172.440 & 31.701 & 0.802 & 0.802 & 0.136 & 0.865 & 0.927 & 1.894 \\
Cat-4 & 1.424 & 297.295 & 8.573 & 164.344 & 53.466 & 0.936 & 0.922 & 0.175 & 0.671 & 0.739 & 1.531 \\
Cat-5 & 1.309 & 289.079 & 12.108 & 176.399 & 32.103 & 0.564 & 0.510 & 0.069 & 0.552 & 0.815 & 1.608 \\
Cat-6 & 1.363 & 292.946 & 10.196 & 171.647 & 33.632 & 1.161 & 1.295 & 0.153 & 0.508 & 0.528 & 1.295 \\
Cat-7 & 1.092 & 272.099 & 1.757 & 134.148 & 157.438 & 0.070 & 0.032 & 0.735 & 0.151 & 0.207 & 1.477 \\
Cat-8 & 1.702 & 315.466 & 13.507 & 198.564 & 40.819 & 1.516 & 2.235 & 0.078 & 0.605 & 0.787 & 1.540 \\
Cat-9 & 1.322 & 289.987 & 12.401 & 177.414 & 30.336 & 1.255 & 1.236 & 0.094 & 0.513 & 0.762 & 2.080 \\
Cat-10 & 1.323 & 290.119 & 11.714 & 174.989 & 39.721 & 1.644 & 2.010 & 0.103 & 0.559 & 0.703 & 1.775 \\
Cat-11 & 1.179 & 279.187 & 12.522 & 172.723 & 53.187 & 1.059 & 4.368 & 0.215 & 0.597 & 0.867 & 1.135 \\
Cat-12 & 1.763 & 319.209 & 11.402 & 191.259 & 52.717 & 1.336 & 9.616 & 0.073 & 0.584 & 0.645 & 1.162 \\
Cat-13 & 1.164 & 277.938 & 12.850 & 171.222 & 33.757 & 1.161 & 11.092 & 0.090 & 0.578 & 0.645 & 1.566 \\
Cat-14 & 0.750 & 240.119 & 9.135 & 137.437 & 26.660 & 1.144 & 4.258 & 0.113 & 0.705 & 0.859 & 2.178 \\
Cat-15 & 1.505 & 302.787 & 8.983 & 174.124 & 37.043 & 1.144 & 3.165 & 0.126 & 0.849 & 0.877 & 1.119 \\
Cat-16 & 0.982 & 262.608 & 11.737 & 155.362 & 28.768 & 1.315 & 3.165 & 0.106 & 0.618 & 0.792 & 0.671 \\
Cat-17 & 1.319 & 289.800 & 12.765 & 179.056 & 38.329 & 1.846 & 1.976 & 0.093 & 0.664 & 0.881 & 1.299 \\
Cat-18 & 1.407 & 296.099 & 7.887 & 163.920 & 57.217 & 0.493 & 0.435 & 0.159 & 0.676 & 0.816 & 0.397 \\
Cat-19 & 1.174 & 278.770 & 10.468 & 164.726 & 29.112 & 1.541 & 2.118 & 0.169 & 0.664 & 0.937 & 1.712 \\
Cat-20 & 0.763 & 241.484 & 13.324 & 149.672 & 30.417 & 1.492 & 5.427 & 0.099 & 0.601 & 0.733 & 1.311 \\
Cat-21 & 1.881 & 326.206 & 10.618 & 190.683 & 50.954 & 1.126 & 1.198 & 0.118 & 0.482 & 0.611 & 1.453 \\
Cat-22 & 1.495 & 302.116 & 10.666 & 180.647 & 35.860 & 0.841 & 29.488 & 0.080 & 0.512 & 0.694 & 1.744 \\
Cat-23 & 1.607 & 309.524 & 12.489 & 190.705 & 32.421 & 1.161 & 9.616 & 0.094 & 0.607 & 0.784 & 1.207 \\
Cat-24 & 1.334 & 290.866 & 11.378 & 176.911 & 36.800 & 1.144 & 3.608 & 0.090 & 0.689 & 0.734 & 1.102 \\
\hline
 Mean* & 1.368 & 291.791 & 10.903 & 173.167 & 38.886 & 1.144 & 4.410 & 0.121 & 0.634 & 0.771 & 1.402 \\
 $\pm$1$\sigma$ & 0.285 & 21.610 & 1.761 & 13.441 & 9.530 & 0.329 & 6.112 & 0.041 & 0.103 & 0.102 & 0.409
\enddata

\tablenotetext{a}{Concentration defined by ratio of the virial radius and the scale radius; $R_{\mathrm{vir}}/R_{\mathrm{s}}$.}

\tablenotetext{b}{The radius at which the \Vmax occurs.}

\tablenotetext{c}{Redshift of host formation defined as when the host main branch progenitor mass equals 0.5\Mvirzero.}

\tablenotetext{d}{Redshift of last major merger defined as a halo with \nicefrac{1}{3} mass merging into the main branch of the host.}

\tablenotetext{e}{Fraction of the host mass in subhalos.}

\tablenotetext{f}{Distance to the closest contamination particle from the host.}

\tablenotetext{*}{Means and deviations were calculated over all halos \textit{except} Cat-7 as it has undergone a very recent major merger.}

\end{deluxetable*}

\begin{figure}[h]
\includegraphics[width=0.48\textwidth]{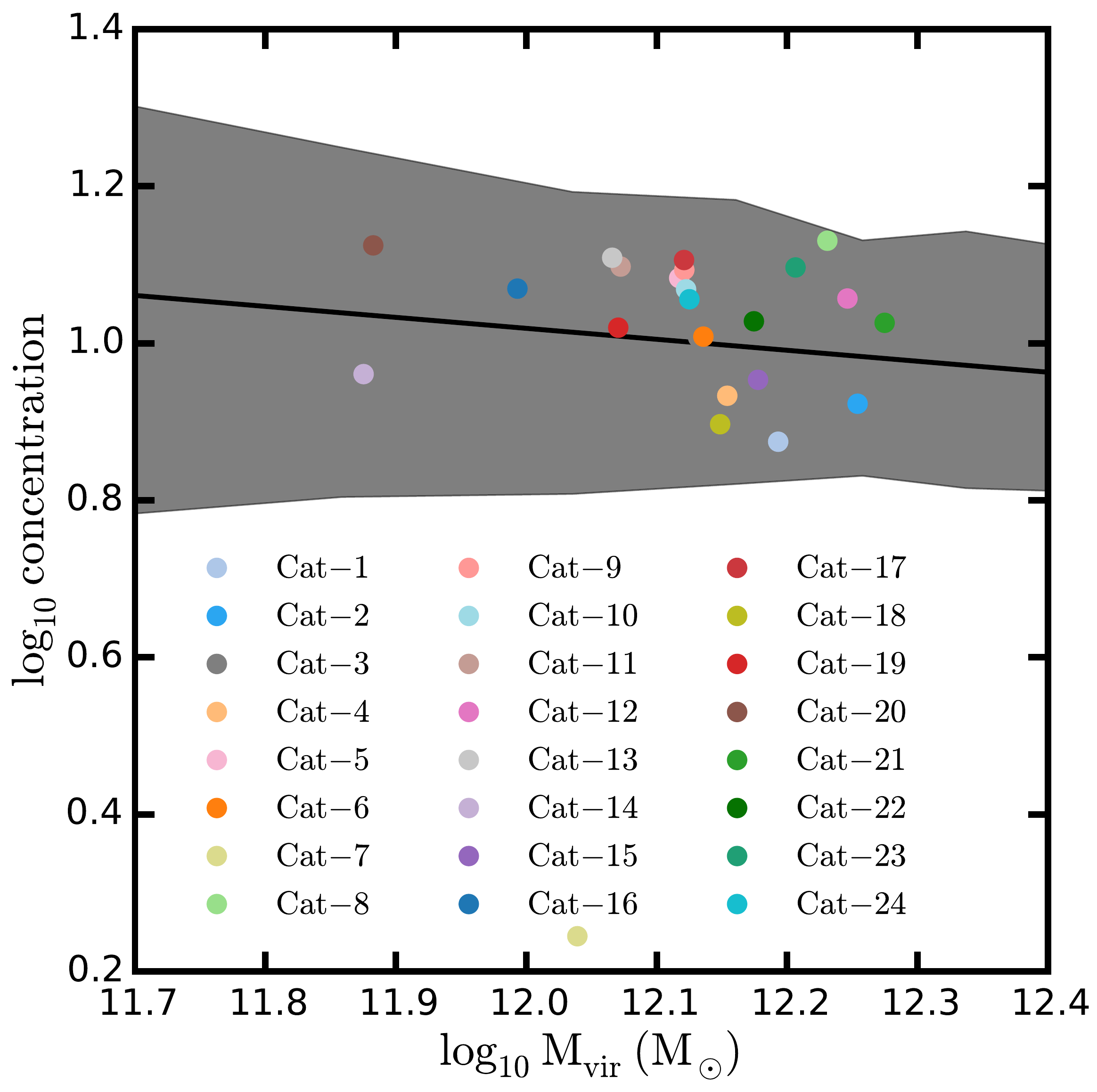} 
\caption{Concentration-mass relation for the 24 $Caterpillar$ halos relative to those found in the parent simulation of similar mass. The concentration is defined as $r_{vir}/r_s$. Solid circles are the zoom-in simulations and the black line represents the concentration-mass relation drawn from the parent simulation for relaxed halos. The grey band is the 1$\sigma$ dispersion in the c--M relation for halos in the parent simulation between ${\rm11.5 <  log_{10}\ M_\odot < 12.5}$.}
\label{fig:concentrations}
\end{figure}

In Figure \ref{fig:concentrations} we plot the concentration-mass (c-M) relation of the parent simulation for similarly sized halos (${\rm11.5 <  log_{10}\ M_\odot < 12.5}$, grey band indicating the 1$\sigma$ dispersion) and overlay the concentration (\Rvir/\Rs) and host mass of the high resolution halos. This shows that for nearly all of the halos, we are sampling within 68$\%$ of the average c-M relation at a fix halo mass. Again, Cat-7 is an outlier with an extremely low concentration because it recently underwent a major merger and has an extremely large substructure mass fraction so its concentration is not meaningful. For this reason we do not include it in the quantitative analysis in terms of determining average halo profile shapes or the mass function slopes. Its properties are still shown and plotted in the various tables and figures, however. Recently, \cite{2015ApJ...809...49B} found that the thickness of planes of satellites depends on the concentration of the host halo. Specifically, they found the thinnest planes are only found in the most concentrated, and hence earliest formed halos. The fact that we sample relatively average concentrations for halos of this mass range means that it is less likely that these hosts will contain planes of satellites, or if they do, their thicknesses will be quite large (Ji et al., in prep.). As the $Caterpillar$ sample grows, we will eventually sample overly concentrated halos, enabling us to see in better detail how this concentration-plane relation holds.

\subsection{Visualizing The Halos $\&$ Their Assembly Histories}

\begin{figure*}[h!]
\centering
\includegraphics[width=1\textwidth]{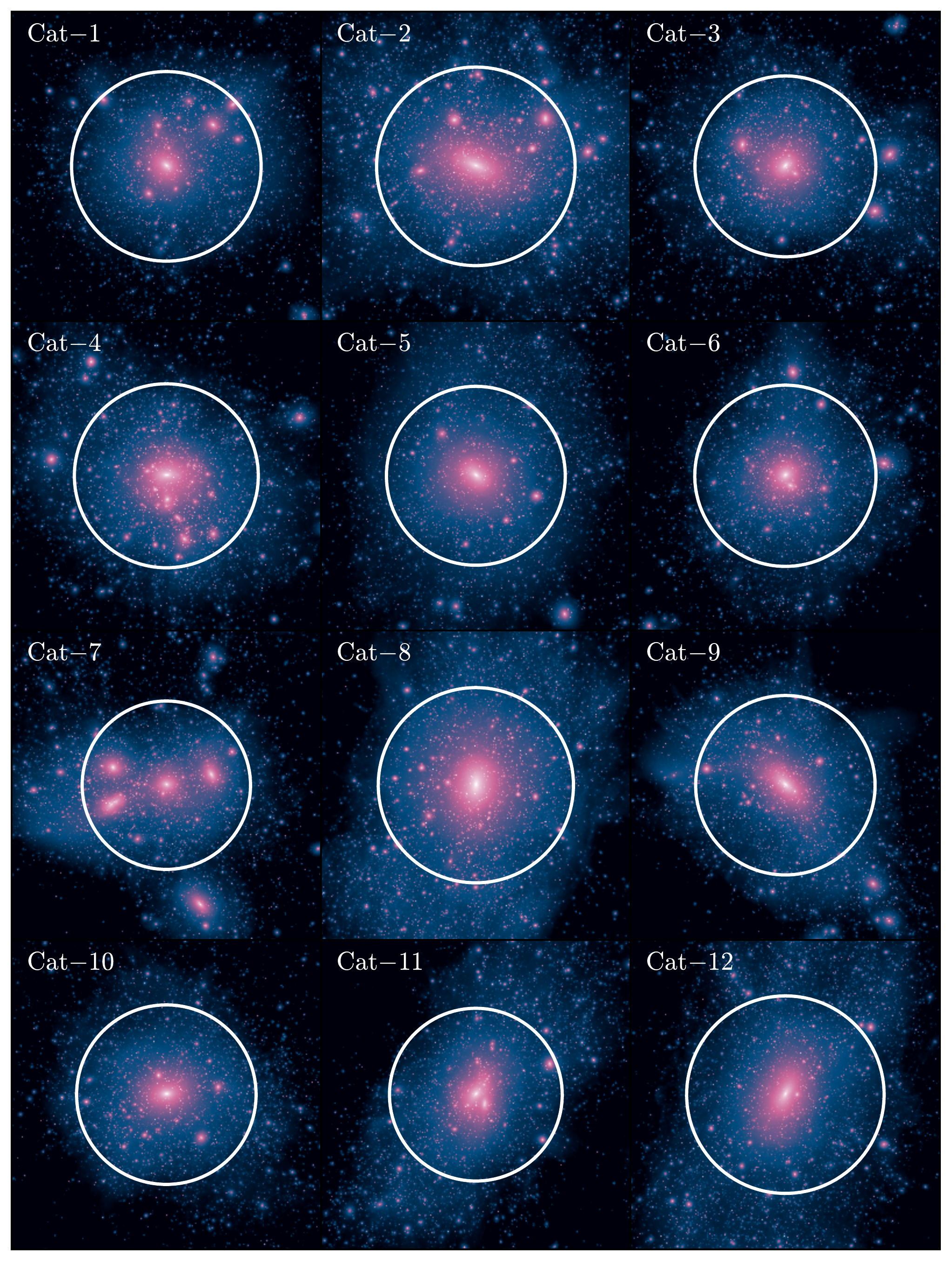}
\caption{Projected dark matter density at $z$ = 0 of the first 12 $Caterpillar$ halos with a box width of 1 Mpc. The image brightness is proportional to the logarithm of the dark matter density squared. Higher resolution images and animations are available at \url{www.caterpillarproject.org}.}
\label{fig:dmprojections1}
\end{figure*}

\begin{figure*}
\centering
\includegraphics[width=1\textwidth]{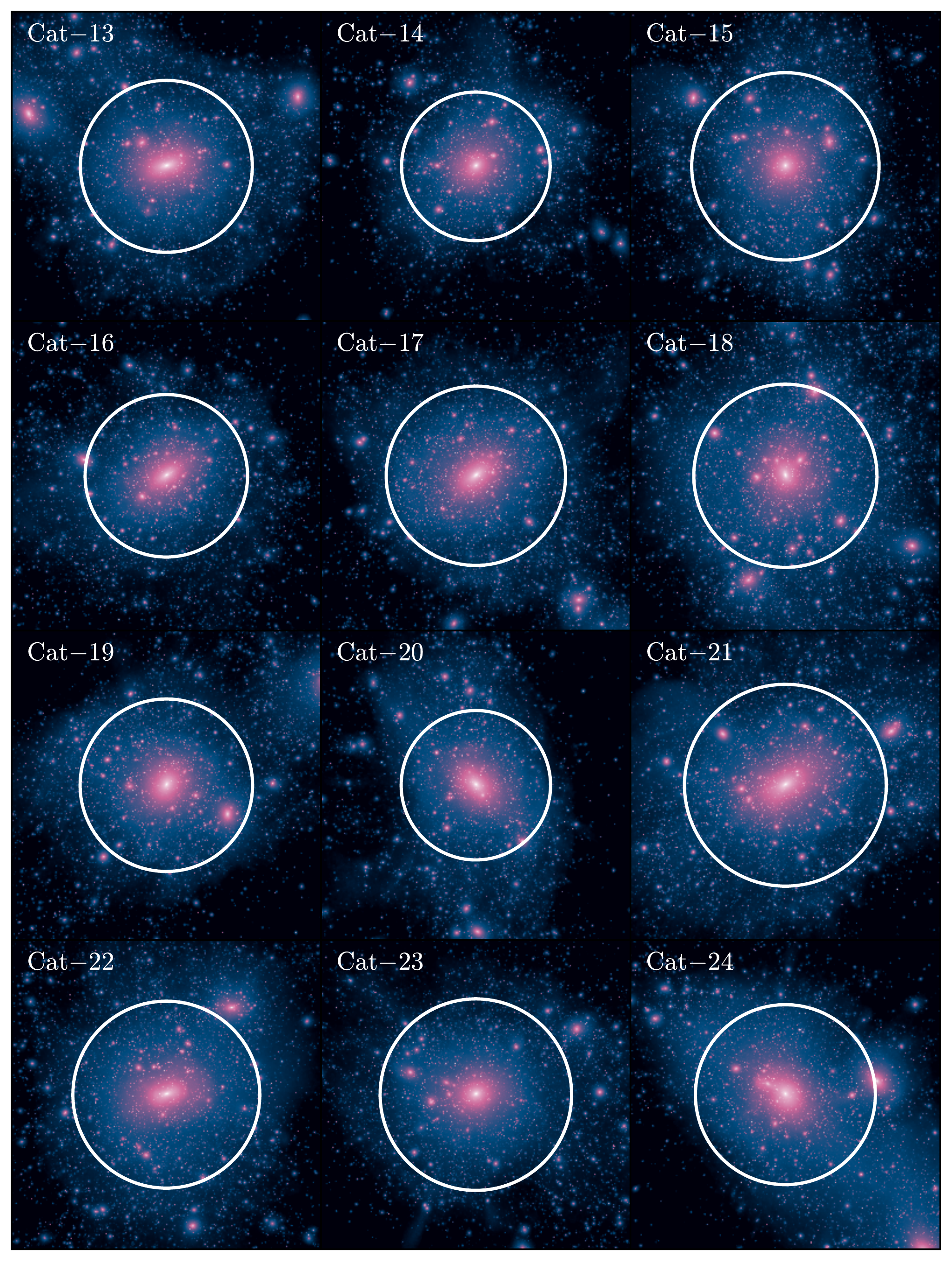}
\caption{Projected dark matter density at $z$ = 0 of each of the second set of 12 $Caterpilar$ halos with a box width of 1 Mpc. The image brightness is proportional to the logarithm of the dark matter density squared. Higher resolution images and animations are available at \url{www.caterpillarproject.org}.}
\label{fig:dmprojections2}
\end{figure*}

In Figures \ref{fig:dmprojections1} and \ref{fig:dmprojections2} we show images of the dark matter distribution in each of our 24 high-resolution halos at redshift $z = 0$. The brightness of each pixel is proportional to the logarithm of the dark matter density squared (i.e.\ log($\rho^2$)projected along the line of sight. To enhance the density contrast, each panel has a different maximum density. We note that similarly colored pixel in one panel does not necessarily mean the density is the same for another panel. The panel width is 1 Mpc and the local dark matter density of the particles in each pixel is estimated with an SPH kernel interpolation scheme based on the 64 nearest neighbor particles. Upon first inspection it is clear that each halo is littered with an abundance of dark matter substructures of varied shapes and sizes. In some cases, there are reasonably large neighbors (e.g.,\ Cat--4, 7, 11, 24). By virtue of our selection criteria these neighbors are no larger than 0.5$\times$\Mvir of the central host. In any case, in under a Gyr, these SMC/LMC sized systems (M$_{peak} > 10^{11}$ \Msol) will likely undergo a major merger with the host galaxy.

\begin{figure}
\includegraphics[width=0.48\textwidth]{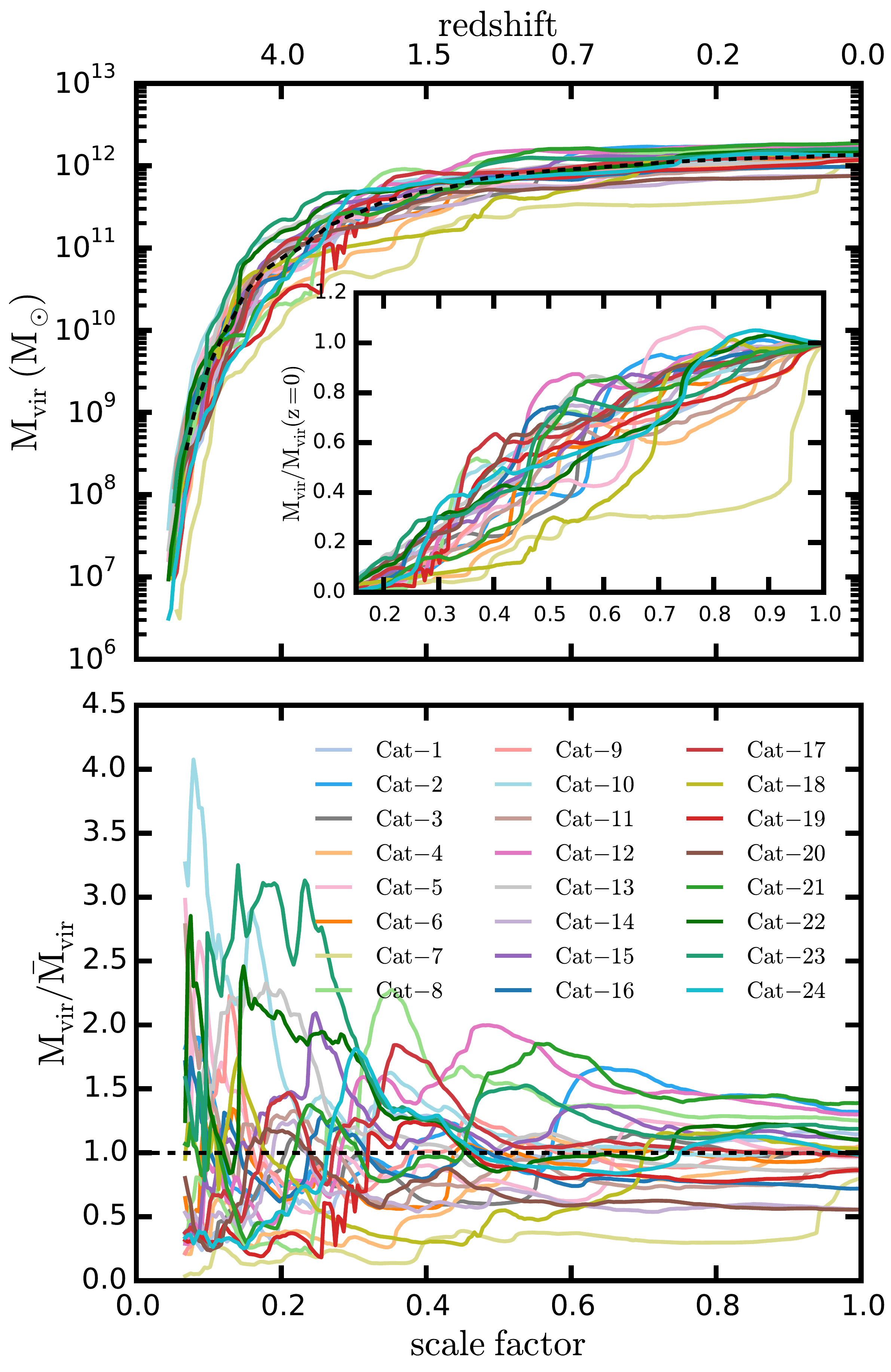} 
\caption{Mass evolution of the first 24 \textit{Caterpillar} halos. The top panel shows each halo evolution along with the mean (black dashed) evolution for each of the halos. The inset panel shows the mass evolution normalized to the halo mass at $z = 0$. In the lower panel we show the mass evolution divided by the mean evolution to enhance individual features for each halo. We sample a diverse assembly history, from extremely quiet through to extremely violent (redshift of last major merger, $z$ $\sim$ 0.07).}
\label{fig:massevolution}
\end{figure}

In Figure \ref{fig:massevolution} we show the mass evolution of each of the halos. As highlighted by the inset which shows the normalized mass evolution, there is a wide variety of formation histories. In our initial catalogue of 24 halos, six halos (Cat--2, 3, 4, 5, 7, 18) have had major mergers since $z = 1$. The halos going above a normalized mass ratio of 1.0 have had a halo pass through them relatively recently which momentarily gives them extra mass such that it is larger than their $z$ = 0 mass (e.g,\ Cat--2).  This indicates that many of the halos are yet to reach an equilibrium state.

Adopting the same criteria as \cite{Neto:2007fp} we assess whether the host halos are relaxed. If their substructure mass fraction is below 0.1, their normalized offset between the center of mass of the halo (i.e.,\ computed using all particles within \Rvir) and the potential center ($\mathrm{x_{off} \equiv |r_c-r_{cm}|/R_{vir}}$, ${\rm r_c\ \equiv}$ center of the potential well, ${\rm r_{cm}\ \equiv}$ center of mass) is below 0.07 and their virial ratio ($2T/|U|$) is below 1.35, then the host is considered relaxed. In Table \ref{tab:prof} we provide the relaxed state of the halo. Many of the halos are in fact unrelaxed under this definition which is by design -- we are sampling a wide range of assembly histories and so halos with recent merger events that prevent the halos from being fully virialized naturally make up part of our sample.

\subsection{Host Halo Profiles}

\begin{figure}
\includegraphics[width=0.48\textwidth]{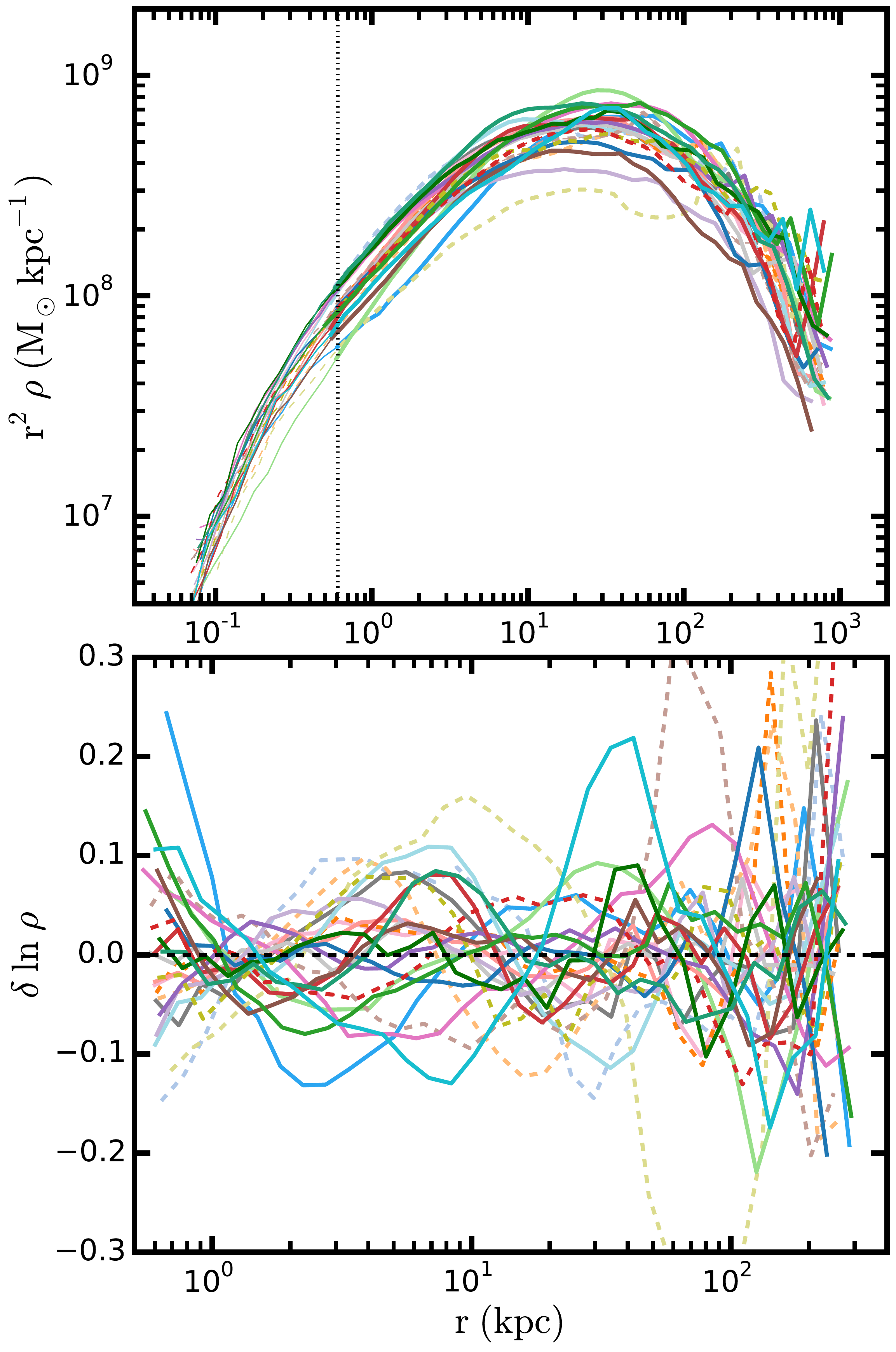} 
\caption{\textit{Upper panel}: Normalized halo profiles for each of the host halos. Relaxed (solid lines) halos are better fit by Einasto profiles ($\alpha \sim 0.16$) over NFW profiles though for halos that are unrelaxed (dashed lines), they are unable to be properly fit by NFW or Einasto profiles which is in agreement with \cite{Neto:2007fp} and \cite{Navarro:2010hna}. This inability stems purely from their definition, which \textit{a priori} assumes that the halos are in virial equilibrium which is clearly not the case for halos which have undergone a recent merger. The dashed black line indicates the \cite{Power03} resolution limit as set by our softening length (also represented by a thinner density profile line). \textit{Lower panel}: The log of the ratio between each of the Einasto fits and the data between the \cite{Power03} radius and the virial radius.}
\label{fig:haloprof}
\end{figure}

In Figure \ref{fig:haloprof} we plot the spherically averaged halo profiles for each of our 24 simulated halos. We draw the measured density profile as a thick line of a given color and continue the fit beyond the smallest radius possible set by \cite{Power03} as a vertical black dashed line. We truncate each fit at this radius. There is a clear diversity in the profile shapes owing in part to the assembly histories of each halo. Halos which have undergone a recent major merger whose substructure mass fractions are higher than average are primarily dominated by a single subhalo (e.g,\ Cat--7 has a large subhalo at 200 kpc). The fitting formula we have used to describe the mass profile of our simulated halos follow the method of \cite{Navarro:2010hna} and is given by the following Einasto form (over all particles within the virial radius):

\begin{equation}
{\rm log[\rho(r)/\rho_{-2}] = (-2/\alpha)\left[(r/r_{-2})^\alpha - 1\right]}.
\end{equation}

The $r_{-2}$ is the scale length of the halo which can be obtained without resorting to a particular fitting formula. We compute the logarithm of the slope profile and identify where a low-order polynomial fit to it intersects the isothermal value ($\gamma = 2$). Unlike the Navarro-Frenk-White profile (NFW) the peak parameter in the Einasto profile, $\alpha$, is allowed to vary and thus provides a third parameter for the fitting formula. The best fitting parameters are found by minimizing the deviation between model and simulation at each bin. Specifically we minimize the function $Q^2$, defined as:

\begin{equation}
{\rm Q^2 = \frac{1}{N_{bins}} \sum\limits_{i=1}^{N_{bins}} (ln(\rho_i) - ln(\rho^{model}_i))^2.}
\end{equation}

In this manner we find a function which clearly illustrates the deviation of the simulated and model profiles. In Table \ref{tab:prof} we show our minimum $Q^2$ parameter ($Q_{min}$), characteristic scale radius $r_{-2}$, and their corresponding densities for each halo. For our relaxed halos, $Q_{min}$ for our Einasto fits are 0.027 $\pm$ 0.010 indicating reasonable agreement between the simulated and model Einasto profiles. This is better than our NFW profile fits for which we obtain $Q_{min} = 0.055 \pm 0.020$. The peak parameters for our Einasto fits are 0.169 $\pm$ 0.023 which is comparable to those of the $Aquarius$ halos ($\alpha$ = 0.145 -- 0.173) studied in \cite{Navarro:2010hna}. For halos which significantly deviate from the mean, it is important to note that those halos are not relaxed and so by definition will not provide meaningful Einasto/NFW fits. A more detailed study of the halo density profiles are reserved for future work.

\begin{deluxetable}{ccccccc}
\tablecolumns{6}
\tablewidth{0pt} 
\tablecaption{The relaxed nature and Einasto profile parameters for the first 24 $Caterpillar$ halos. For comparison, $Q_{min}$ fits for NFW halo profiles are also listed. \label{tab:prof}}
\tablehead{  
 \colhead{Halo} &  
 \colhead{Relaxed\tablenotemark{$a$}} & 
 \colhead{$\rho_{-2}$\tablenotemark{$c$}} & 
  \colhead{$r_{-2}$\tablenotemark{$d$}} &
 \colhead{$\alpha$\tablenotemark{$b$}}  & 
  \colhead{$Q_{min}$} & 
  \colhead{$Q_{min}$} \\
 \colhead{Name} & 
 \colhead{} & 
 \colhead{} & 
 \colhead{($\times10^5$)} &  
 \colhead{} & 
 \colhead{Ein.} & 
 \colhead{NFW}}
\startdata
Cat-1 & \xmark & 9.929 & 27.182 & 0.128 & 0.039 & 0.103 \\
Cat-2 & \cmark & 3.531 & 46.846 & 0.185 & 0.040 & 0.033 \\
Cat-3 & \cmark & 11.604 & 25.309 & 0.151 & 0.028 & 0.067 \\
Cat-4 & \xmark & 5.504 & 34.962 & 0.154 & 0.039 & 0.080 \\
Cat-5 & \cmark & 13.305 & 24.152 & 0.167 & 0.018 & 0.050 \\
Cat-6 & \xmark & 9.254 & 28.299 & 0.164 & 0.027 & 0.058 \\
Cat-7 & \xmark & 0.482 & 90.528 & 0.075 & 0.082 & 0.162 \\
Cat-8 & \cmark & 8.027 & 34.555 & 0.236 & 0.033 & 0.036 \\
Cat-9 & \cmark & 12.159 & 25.543 & 0.186 & 0.011 & 0.034 \\
Cat-10 & \cmark & 14.482 & 23.492 & 0.168 & 0.029 & 0.049 \\
Cat-11 & \xmark & 10.617 & 25.450 & 0.173 & 0.049 & 0.071 \\
Cat-12 & \cmark & 8.897 & 31.157 & 0.160 & 0.030 & 0.062 \\
Cat-13 & \cmark & 11.842 & 25.010 & 0.187 & 0.009 & 0.036 \\
Cat-14 & \cmark & 10.455 & 21.347 & 0.139 & 0.018 & 0.078 \\
Cat-15 & \cmark & 8.905 & 28.744 & 0.139 & 0.025 & 0.078 \\
Cat-16 & \cmark & 10.445 & 24.162 & 0.166 & 0.026 & 0.056 \\
Cat-17 & \cmark & 11.283 & 26.799 & 0.195 & 0.019 & 0.026 \\
Cat-18 & \xmark & 6.998 & 31.012 & 0.150 & 0.023 & 0.082 \\
Cat-19 & \xmark & 9.079 & 26.881 & 0.164 & 0.032 & 0.054 \\
Cat-20 & \cmark & 11.415 & 22.158 & 0.199 & 0.018 & 0.020 \\
Cat-21 & \cmark & 6.682 & 36.486 & 0.175 & 0.025 & 0.043 \\
Cat-22 & \cmark & 10.857 & 26.957 & 0.156 & 0.017 & 0.063 \\
Cat-23 & \cmark & 13.486 & 26.024 & 0.172 & 0.018 & 0.039 \\
Cat-24 & \cmark & 7.040 & 31.734 & 0.181 & 0.044 & 0.050 \\
\hline
 Mean* & - & 9.817 & 28.446 & 0.169 & 0.027 & 0.055 \\
 $\pm$1$\sigma$ & - & 2.6106 & 5.5677 & 0.023 & 0.010 & 0.020
\enddata

\tablenotetext{a}{Relaxed criteria is based on that of \cite{Neto:2007fp}. If the substructure mass fraction is below 0.1, their center of mass displacement ($\mathrm{x_{off} = |r_c-r_{cm}|/R_{vir}}$) is below 0.07 and their virial ratio ($2T/|U|$) is below 1.35, then the host is considered relaxed.}

\tablenotetext{b}{Einasto slope parameter of the form $\mathrm{\rho(r) \propto exp(-Ar^\alpha)}$.}

\tablenotetext{c}{The density at the characteristic radius in units of: $10^{10} h^2$ M$_\sun$ kpc$^{-3}$.}

\tablenotetext{d}{The characteristic radius or `peak' radius of the $r^2\rho$ profile in units of: $h^{-1}$ kpc.}

\tablenotetext{*}{Means and deviations were calculated over all halos \textit{except} Cat-7 as it has undergone a very recent major merger.}

\end{deluxetable}

\subsection{Subhalo Properties}

\begin{figure*}[!h]
\includegraphics[width=\textwidth]{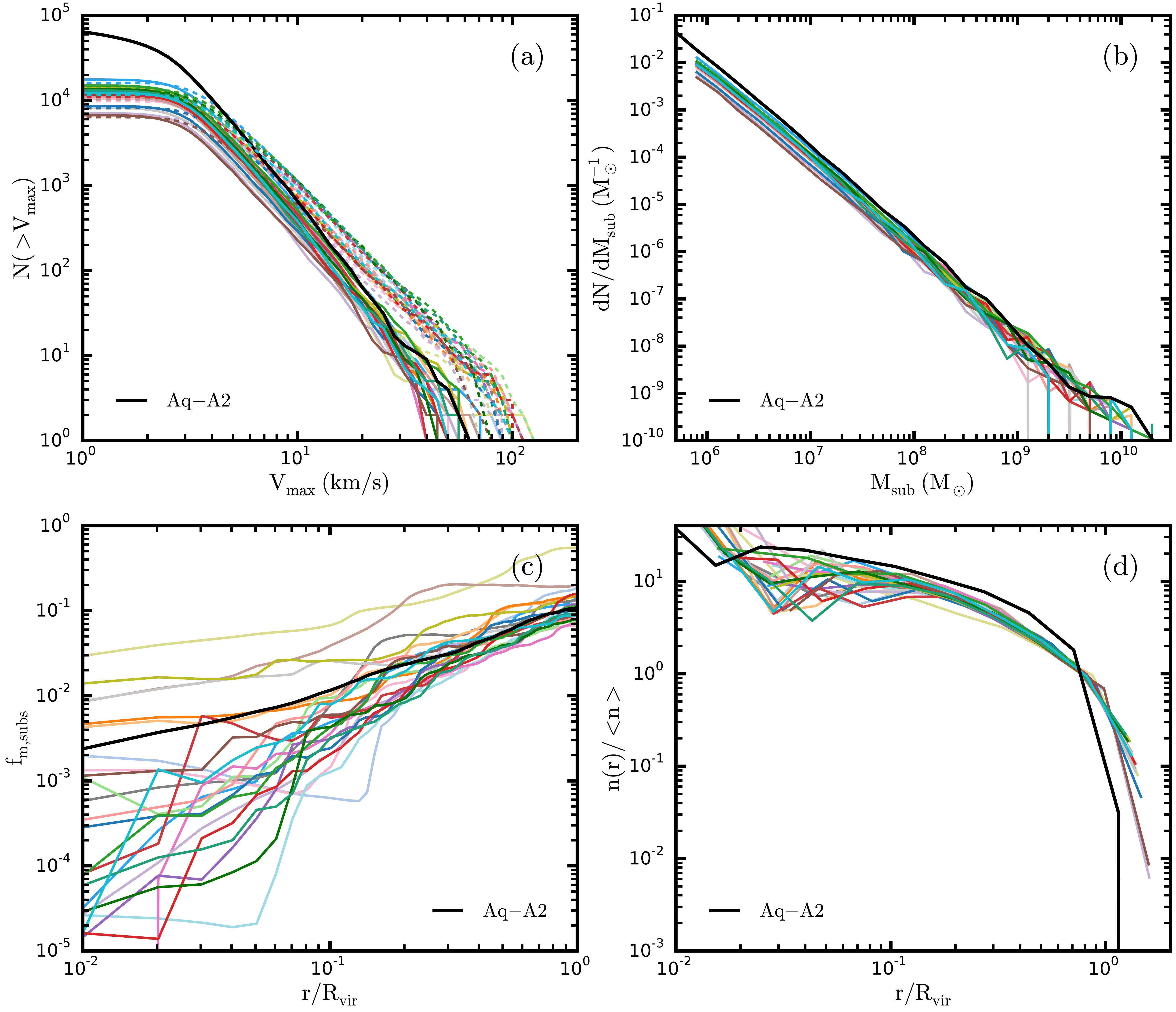} 
\caption{(a) The cumulative abundance of subhalos as a function of their maximum circular velocity (\Vmax) for each of the $Caterpillar$ halos. The solid lines represent subhalos at $z = 0$ and the dashed lines represent halos at infall (i.e.,\ when they first cross the virial radius of the host). Many of the SMC/LMC sized systems (\Mpeak $> 10^{11}$ \Msol) are destroyed by $z$ = 0, though some large subhalos do survive. The $Caterpillar$ suite is complete down to $\sim$4 \kms. See our convergence study in the Appendix A. The solid black line is that of the Aquarius-A halo at a similar resolution (level-2, \citealt{Springel:2008gd}). (b) The subhalo mass functions for each of the host halos. When normalized by mass, the range $10^{-6} \-- 10^{-4}$ \Msol there is small scatter in the subhalo abundances as all of our hosts are very similar in mass.  Over the ranges 10$^{6-8}$ \Msol we obtain a median mass function slope of -1.88 $\pm$ 0.1, slightly shallower than Aquarius ($\alpha \sim 1.90$). As we move to higher and lower mass regimes, there are differing abundances of large and small subhalos for each host. There are a number of systems with $\sim10^{9\--10}$ \Msol halos within the virial radius of the host making them possible Large/Small Magellanic Cloud analogues. Again, the solid black line is that of the Aquarius-A. (c) The subhalo mass fraction as a function of radius scaled by the virial radius of the host. The substructure mass is distributed similarly in nearly all halos with the exception of Cat--7 which has undergone a major merger. There is more variability in the substructure mass fraction at low radii owing to the on-going disruption most prevalent in the inner most dense regions of each host. (d) The normalized number of subhalos as function of radius scaled by the virial radius of the host. The number of subhalos as a function of distance is extremely self-similar across all halos except within the inner 10$\%$ of the virial radius of each host. This is in agreement with the findings of \cite{Springel:2008gd} when factoring in their slightly higher resolution of the $Aquarius$ suite and use of the observationally disfavoured high value of $\sigma_8$ ($\sigma_8$ = 0.9).}
\label{fig:fourpanels}
\end{figure*}

In Figure \ref{fig:fourpanels}a we show the cumulative abundance of subhalos as a function of their maximum circular velocity for each $Caterpillar$ halo. Since we achieve excellent convergence (see Appendix A), we reliably resolve halos with circular velocities of $\sim$4 \kms, which is crucial for identifying the sites of first star formation. At the high \Vmax end we find a variety of different sized subhalos for each host. Some hosts have only one 10 \kms subhalo whereas another host halo has a large 70 \kms subhalo within the virial radius. Between 5-20 \kms all halos are very similar in their \Vmax function slopes within a slight offset owing to normalization stemming from the differences in host halo mass. At low \Vmax values ($\sim$$3$ \kms) we begin to lose completeness of our host subhalo sample due to lack of resolution. We additionally include in this Figure the \Vmax function for subhalos at infall (i.e.,\ when a subhalo first crosses the virial radius of the host). Since dynamical friction affects the highest mass subhalos the fastest, the biggest difference in the functions occurs at the high-mass end whereby several LMC sized systems (\Mpeak $>$ $10^{11}\ M_\odot$) have been destroyed (over a time scale of 1 -- 2 Gyrs) between infall and $z$ = 0. These large LMC sized-systems at infall can host anywhere from 4 -- 30$\%$ of the Milky Way sized halo's subhalos at $z = 0$ depending on their orbit and infall time (Griffen et al. in prep.). In solid black we also plot the $Aquarius$ Aq-A2 halo from \cite{Springel:2008gd} (using the same version of \rockstar that we used for the $Caterpillar$ halos). We find the differences in the cosmology ($\sigma_8=0.9$) and the slightly higher resolution of $Aquarius$ leads to systematic differences in subhalo abundance.

In Figure \ref{fig:fourpanels}b, we show the subhalo mass functions for each of the halos. Our results are best fit by the power law d$N$/d$M\propto M^{-1.88 \pm 0.10}$, which is less steep than that found in the $Aquarius$ halos of \cite{Springel:2008gd}. This slope is the best fit over the ranges $10^{5} \-- 10^{8}$ \Msol. We do observe a scatter in the subhalo abundances. This can be explained by the subtle concentration-subhalo-abundance relation whereby for fixed halo mass, there are more (less) subhalos belonging to hosts which are less (more) concentrated (e.g,\ \citealt{Zentner:2005bq}, \citealt{Watson:2011dq}, \citealt{Mao:2015tx}). Indeed, we find halos which are more concentrated (see Figure \ref{fig:concentrations}) have lower normalizations than those less concentrated at fixed \Vmax. This is simply because halos which are more concentrated have formed earlier and so subhalos have spent substantially longer undergoing dynamical disruption within the host compared to similar sized subhalos orbiting less concentrated hosts. 

In Figure \ref{fig:fourpanels}c we show the subhalo radial mass fraction which indicates high variability in the contribution to the total halo mass from substructure as a function of galactocentric distance. For example, at 0.1\Rvir, the total mass contributing to the host halo mass from substructure varies by a factor of 10 or more when normalized by mass. At \Rvir our substructure mass fraction varies by $\sim10\%$ (see Table \ref{tab:catsuite} for exact fractions). Cat-7 has a large component of the halo mass in substructure at low radii because it has recently undergone a major merger (z = 0.03). Those halos with a large substructure mass fraction generally have had a recent major merger and are in the process of disrupting the recently accreted systems. On average, for a fixed fraction of the virial radius, the $Caterpillar$ halos have less mass in substructure than that found in the Aq-A $Aquarius$ halo (see solid black line, calculated using the exact same code). In Figure \ref{fig:fourpanels}d we plot the normalized number of subhalos as a function of radius scaled by the virial radius of the host. We find the scatter in the number of subhalos as a function of galactocentric distance is a factor of 3 across all halos except within the inner 10$\%$ of the host halo where we are subject to noise in the halo finding produced by \rockstar. Again, in solid black we also plot the $Aquarius$ Aq-A2 halo from \cite{Springel:2008gd}. We find the differences in the cosmology ($\sigma_8=0.9$) and in particular the slightly higher resolution of $Aquarius$ leads to this systematic difference in the subhalo number density.

\subsection{Too Big To Fail}
\label{sec:tbtf}
We also examine halos which are massive enough to form stars but have no luminous counterpart in the nearby Universe (i.e.,\ \textit{the too big to fail} problem, hereafter TBTF, \citealt{BoylanKolchin:2011tn}). To do this we select halos with \Vpeak $>$ 30 \kms which are subhalos large enough to retain substantial gas in the presence of an ionizing background and therefore theoretically should form stars. We follow the same definition as in \cite{GarrisonKimmel:2014bv} to count two classes of halos. Strong massive failures are too dense to host any of the currently known bright MW classical dwarf spherioidals (dSph) galaxies. Massive failures (MFs) include all strong massive failures (SMFs) as well as all massive subhalos which have densities consistent with the high-density dSphs (i.e.,\ Draco and Ursa Minor) but can not be associated with them without allowing a single dwarf galaxy to be hosted by multiple halos (i.e.,\ assuming every observable dSph galaxy is hosted by exactly \textit{one} halo). Most subhalos in the range of \Vmax = 25 -- 30 \kms could host a low-density dwarf and as such are not defined as massive failures. 

\begin{figure*}
\includegraphics[width=1\textwidth]{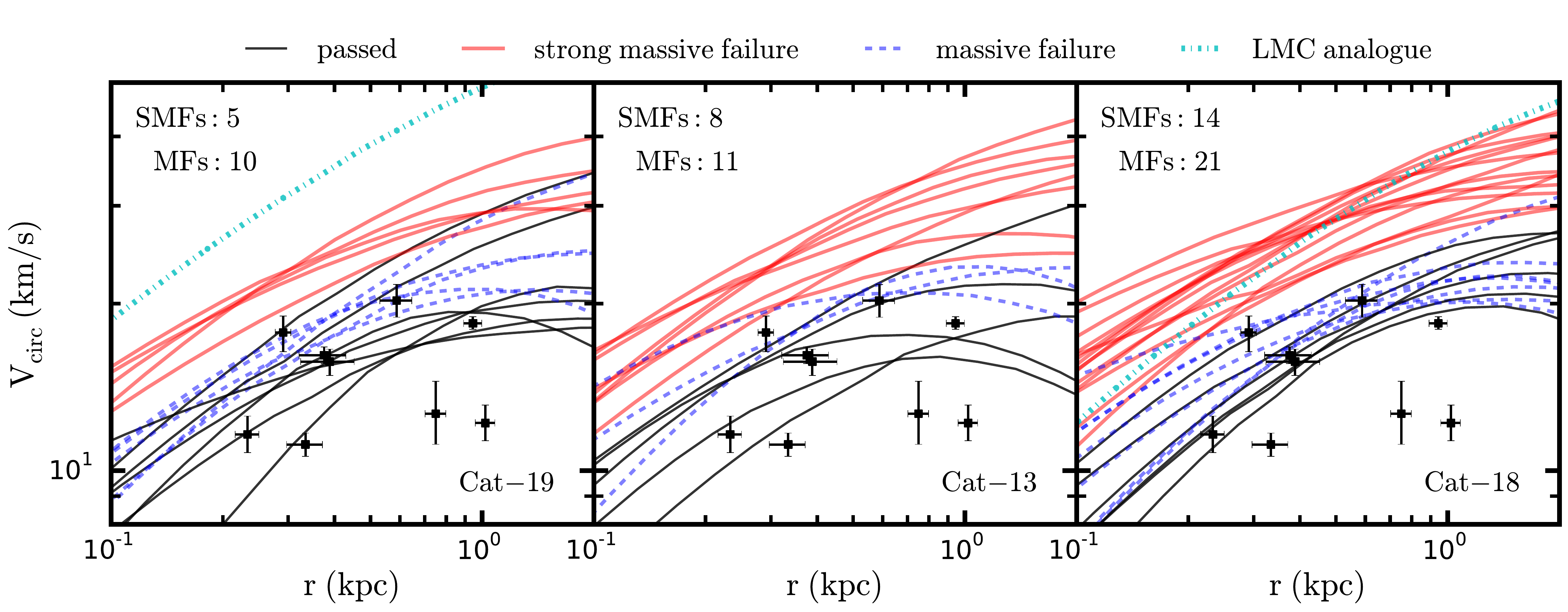} 
\caption{The rotation curves for all subhalos identified with a peak maximum circular velocity \Vpeak, above 30 \kms within 300 kpc of three different $Caterpillar$ hosts. The black squares indicate observational constraints from dwarf galaxies from \cite{Wolf:2010df}. We only include those observed systems whose luminosities are above $2\times10^5$ L$_{\odot,V}$ (i.e.,\ showing only classical dwarfs and excluding ultra-faint dwarf galaxies). Dashed-dot cyan lines are LMC analogues (i.e.,\ \Vmax $>$ 60 \kms), blue dashed lines are massive failures (MFs) and red solid lines are strong massive failures (SMFs). Thin black solid lines are subhalos pass the check and have at least one observed dwarf with a comparable circular velocity (i.e.,\ one halo passes through the circular velocity of a known classical dwarf determined by \citealt{Wolf:2010df}).}
\label{fig:tbtfpanel}
\end{figure*}

In Figure \ref{fig:tbtfpanel}, we plot a sample of the rotation curves for three different $Caterpillar$ halos (Cat-19, Cat-13 and Cat-18). We adopt the \cite{BoylanKolchin:2011vs} Einasto correction for $R < 291$ pc, which differ from the $ELVIS$ profile fits in that they extrapolate their entire profile from \Rmax with various density profile shapes. Black squares depict circular velocities of the classical dwarf galaxies (with luminosities above $2\times10^5$ L$_{\odot,V}$), as measured by \cite{Wolf:2010df}.  Dashed-dot cyan lines are LMC analogues (i.e.,\ \Vmax $>$ 60 \kms which are excluded from our failure analysis), blue dashed lines are massive failures and red solid lines are strong massive failures. Thin black solid lines are subhalos which pass the test of having at least one observed dwarf with a comparable circular velocity, i.e.,\ a circular profile goes through one of the observed dwarf galaxy data points. The cumulative number of profiles above and below the observed classical dwarfs for Cat-19 are 10 MFs, 5 SMFs and one LMC analogue. Similarly we find Cat-13 has 11 MFs and 8 SMFs. Cat-18 has the most failures of any halo with 21 MFs and 14 SMFs with one LMC analogue.

\begin{figure}
\includegraphics[width=0.48\textwidth]{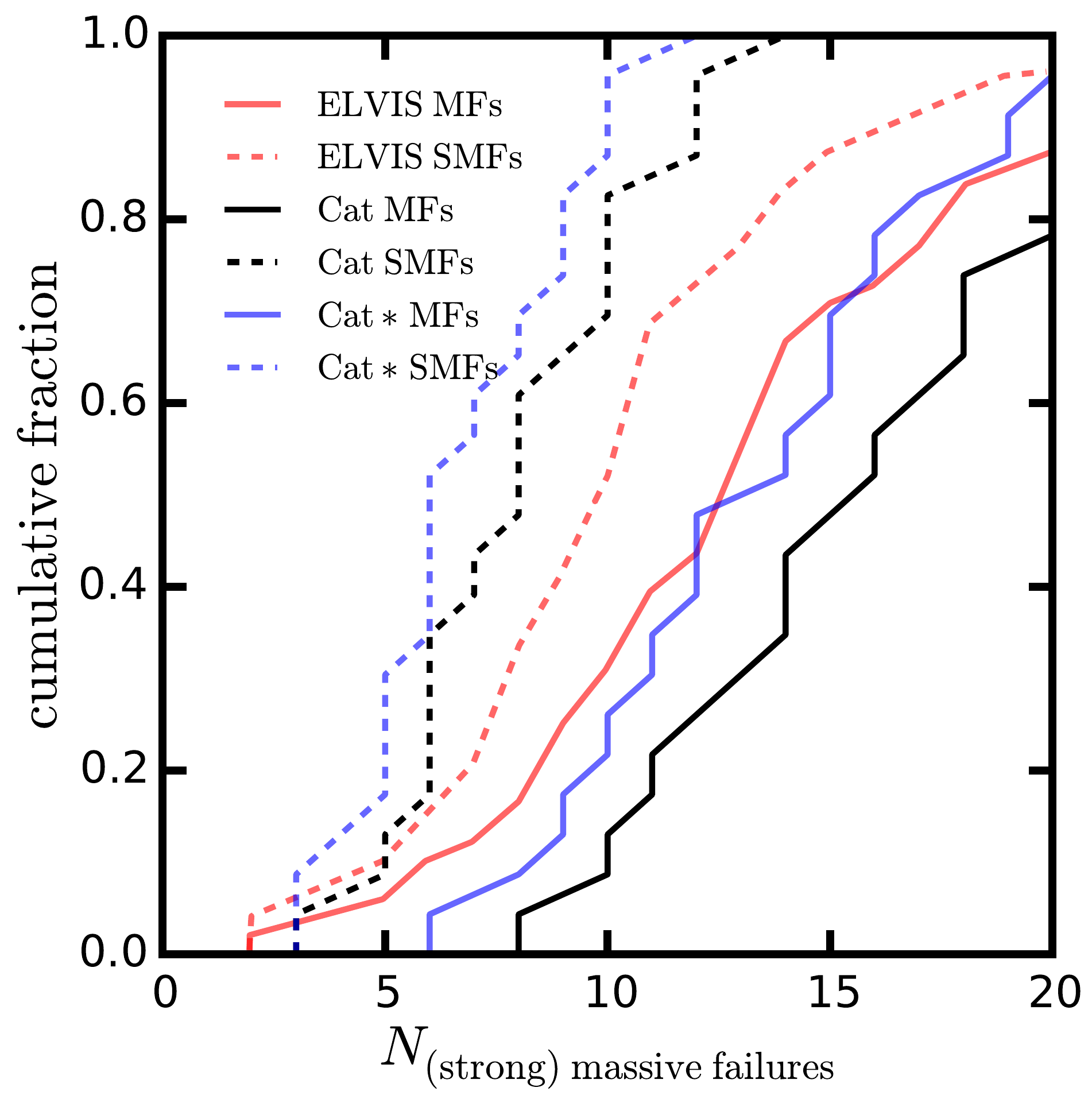} 
\caption{Cumulative fraction of massive failures (MFs) and strong massive failures (SMFs) across all $Caterpillar$ halos  (excluding Cat-7 which has undergone a recent massive major merger). We predict 8 $\pm$ 3 (1$\sigma$) strong massive failures and 16 $\pm$ 5 (1$\sigma$) massive failures within 300 kpc of the Milky Way. We also include the result from \cite{GarrisonKimmel:2014bv} (i.e,\ using $\alpha= 0.15$ Einasto profiles) as a reference. The black lines are all failures which are detected by our version of \rockstar which includes iterative unbinding. The blue lines are all failures found using a cut which mimics \rockstar without iterative unbinding (see Section \ref{sec:tbtf} text for details). This indicates a fraction of failures are undergoing tidal disruption which may have been unaccounted for in the $ELVIS$ subhalo catalogues.}
\label{fig:tbtfcum}
\end{figure}

In Figure  \ref{fig:tbtfcum} we plot the number of strong and massive failures across all Caterpillar halos. Specifically we plot the fraction of hosts with fewer than N MFs and N SMFs within 300 kpc of each host as a function of N (black lines). Averaging over the entire Caterpillar sample (excluding Cat-7 as it has recently had a massive major merger), we predict 8 $\pm$ 3 (1$\sigma$) SMFs and 16 $\pm$ 5 (1$\sigma$) MFs within 300 kpc. If the Milky Way were well described by such an average we would expect to have these failures.

For comparison, we plot the $ELVIS$ MF and SMF counts (red lines). The lower resolution in these simulations requires an extrapolation of the velocity profile from \Vmax and \Rmax using an analytic Einasto profile. Qualitatively, both simulation suites agree that there are a significant number of both MFs and SMFs. Quantitatively, there are several differences, which we now describe. The $Caterpillar$ suite has many more MFs. This is due to our ability to better resolve high \Vpeak subhalos that have been tidally stripped. In particular, the iterative unbinding procedure described in Section \ref{sec:updaterockstar} removes the need for the \rockstar \unboundthreshold parameter \citep{Behroozi:2013cn}. We can simulate the effect of the standard \unboundthreshold = 0.5 cut by removing halos whose bound mass is less than 50$\%$ of their mass prior to unbinding. The MF counts with this cut are shown in Figure  \ref{fig:tbtfcum} by the blue lines, which are very similar to the $ELVIS$ MF counts.

The $Caterpillar$ suite also has significantly fewer SMFs compared to $ELVIS$. This discrepancy is likely due to the fact that we have measured rather than extrapolated the subhalo density profiles. Variations in the Einasto shape parameter ($\alpha$) greatly affect the massive failure count (see Figure 4 of \citealt{GarrisonKimmel:2014bv}). The Einasto fits to our density profiles have $\alpha$ typically closer to 0.2, which is less discrepant.

Whilst TBTF is a prevalent problem in pure $N$-body simulations, many authors have indicated the tension between the circular velocities of observed classical dwarfs and simulated subhalos can be alleviated with the addition of supernovae feedback and ram pressure stripping (e.g.,\ \citealt{Pontzen:2012jg}, \citealt{Zolotov:2012hi}, \citealt{Arraki:2014ku}, \citealt{Brooks:2013fb}, \citealt{DelPopolo:2014kf}, \citealt{Gritschneder:2013km}, \citealt{Elbert:2015ee}, \citealt{Maxwell:2015fr}) or by making dark matter self-interacting (e.g.,\ \citealt{Vogelsberger:2012dy}, \citealt{Zavala:2013iq}). Our results are within 1$\sigma$ of the number of failures found by \cite{GarrisonKimmel:2014bv} (i.e.,\ 12 massive failures within 300 kpc), even when using a better density profile estimation.

\section{Conclusion}
\label{sec:conclusion}

In this work we have presented the first results of the $Caterpillar$ simulation project, whose goal is to better understand the formation of Milky Way-sized galaxies and their satellite companions at both high and low redshift. We have carried out 24 initial simulations in a $Planck$ based $\Lambda$CDM cosmology. Although the total halo number will increase to 60 -- 70 shortly, these first 24 halos provide us with an exquisite initial set of data to achieve our first set of science goals. In our approach, we have taken exceptional care to validate our numerical techniques. We quadruple the current number of halos available in the literature at this extremely high mass and temporal resolution, allowing for detailed statistical studies of the assembly of Milky Way-sized galaxies. We additionally have adjusted our simulation parameters to be more inclusive of potential scientific questions not yet studied in simulations of this size (i.e.,\ decreasing the temporal resolution to $\sim$5 Myrs/snapshot and increasing the volume resolved by high-resolution particles to 1--2 Mpc). The results presented above demonstrate our data quality and give initial clues at how halo properties vary across large numbers of realizations. Our initial key results can be summarized as follows:

\begin{enumerate}
\item Key halo properties such as the halo profile, mass functions and substructure fractions are intimately connected to each halo's overall assembly history. Halos which have undergone recent major mergers have profiles which are poorly fit by either the NFW or Einasto profile. For those halos which are well fit by Einasto profiles, they have peak $\alpha$ values of 0.169 $\pm$ 0.023. Excluding the Cat-7 halo, we find a $Q_{min} = 0.027 \pm 0.010$ indicating reasonable agreement with Einasto fits of the $Aquarius$ halos.

\item The abundance of dark matter subhalos remains relatively similar across our sample when normalized to host halo. As such, our halo mass functions are best fit by a simple power law, d$N$/d$M\propto M^{-1.88 \pm 0.10}$. The scatter in the normalizations of the mass functions is due to the concentration-subhalo abundance relation for fixed halo mass (i.e.,\ our more concentrated halos exhibit lower normalizations for fixed $\mathrm{M}_{host}$).

\item Regarding TBTF, dividing halos into two categories of massive failures and strong massive failures we predict 8 $\pm$ 3 (1$\sigma$) strong massive failures and 16 $\pm$ 5 (1$\sigma$) massive failures within 300 kpc of the Milky Way.

\item Iterative unbinding in {\sc{rockstar}} must be included to properly recover all bound subhalos this resolution. We recover 52 halos above 10$^8$ \Msol across a sample of 13 $Caterpillar$ halos ($\sim$4 per host halo) using iterative unbinding which would have otherwise been unaccounted for using traditional {\sc{rockstar}}. This means that a small fraction of massive subhalos undergoing heavy tidal disruption may be unaccounted for in studies using traditional rockstar (e.g.,\ the $ELVIS$ halo catalogues).

\end{enumerate}

This paper outlines the data products of the $Caterpillar$ simulations and sets the foundation of many upcoming in-depth studies of the Local Group. Through our statistical approach to the assembly of Milky Way-sized halos we will gain a more fundamental insight into the origin and formation of the Galaxy, its similar sized cousins and their respective satellites.

\acknowledgements{
BG would like to thank Paul Hsi for assistance with the compute cluster at MKI. He would also like to thank Phillip Zukin and Paul Torrey for helpful discussions. The authors thank Oliver Hahn for making the initial conditions code, {\sc{music}}, publicly available. The authors also thank Volker Springel for making {\sc{gadget-2}} publicly available and for providing a version of {\sc{gadget-3}}/{\sc{gadget-4}} for our use. The authors thank Peter Behroozi for making {\sc{rockstar}} and {\sc{consistent-trees}} publicly available and additionally thank him for technical support in modifying {\sc{rockstar}}.

Support for this work was provided by XSEDE through the grants (TG-AST120022, TG-AST110038). BG and AF acknowledges support of the compute cluster of the Astrophysics Division which was built with support from the Kavli Investment Fund administered by the MIT Kavli Institute for Astrophysics and Space Research. GD acknowledges support by NSF Grant 1122374. BWO and FG were supported through the NSF Office of Cyberinfrastructure by grant PHY-0941373 and by the Michigan State University Institute for Cyber-Enabled Research (ICER).  BWO was supported in part by by NSF grant PHY 08-22648 (Physics Frontiers Center/Joint Institute for Nuclear Astrophysics) and NSF Grant  PHY-1430152 (JINA Center for the Evolution of the Elements). AF acknowledges support from the Silverman (1968) Family Career Development professorship.
}

\section*{Appendix A: Convergence Study}
\label{sec:AppendixA}

\begin{figure}
\includegraphics[width=0.49\textwidth]{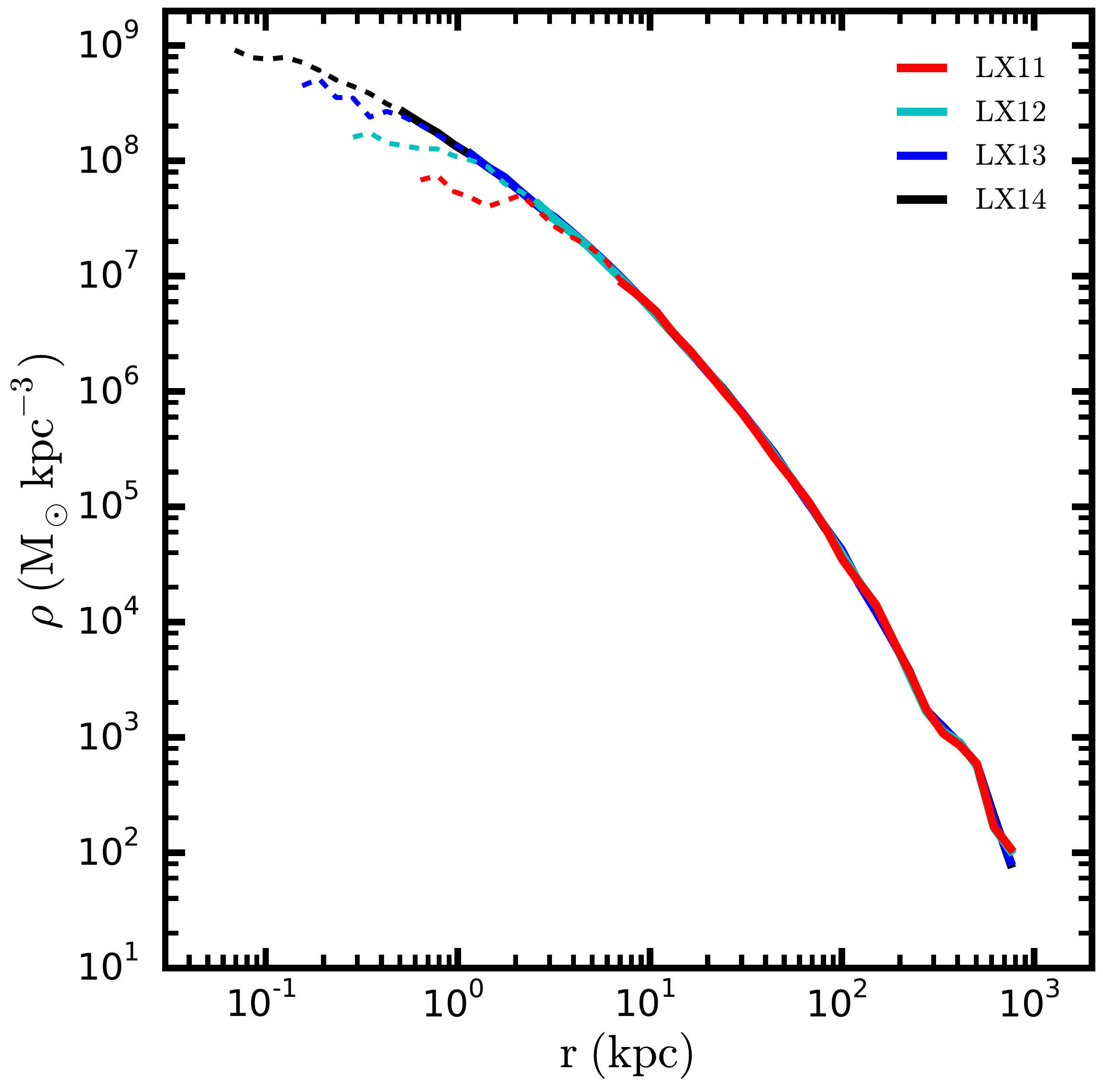}  
\includegraphics[width=0.5\textwidth]{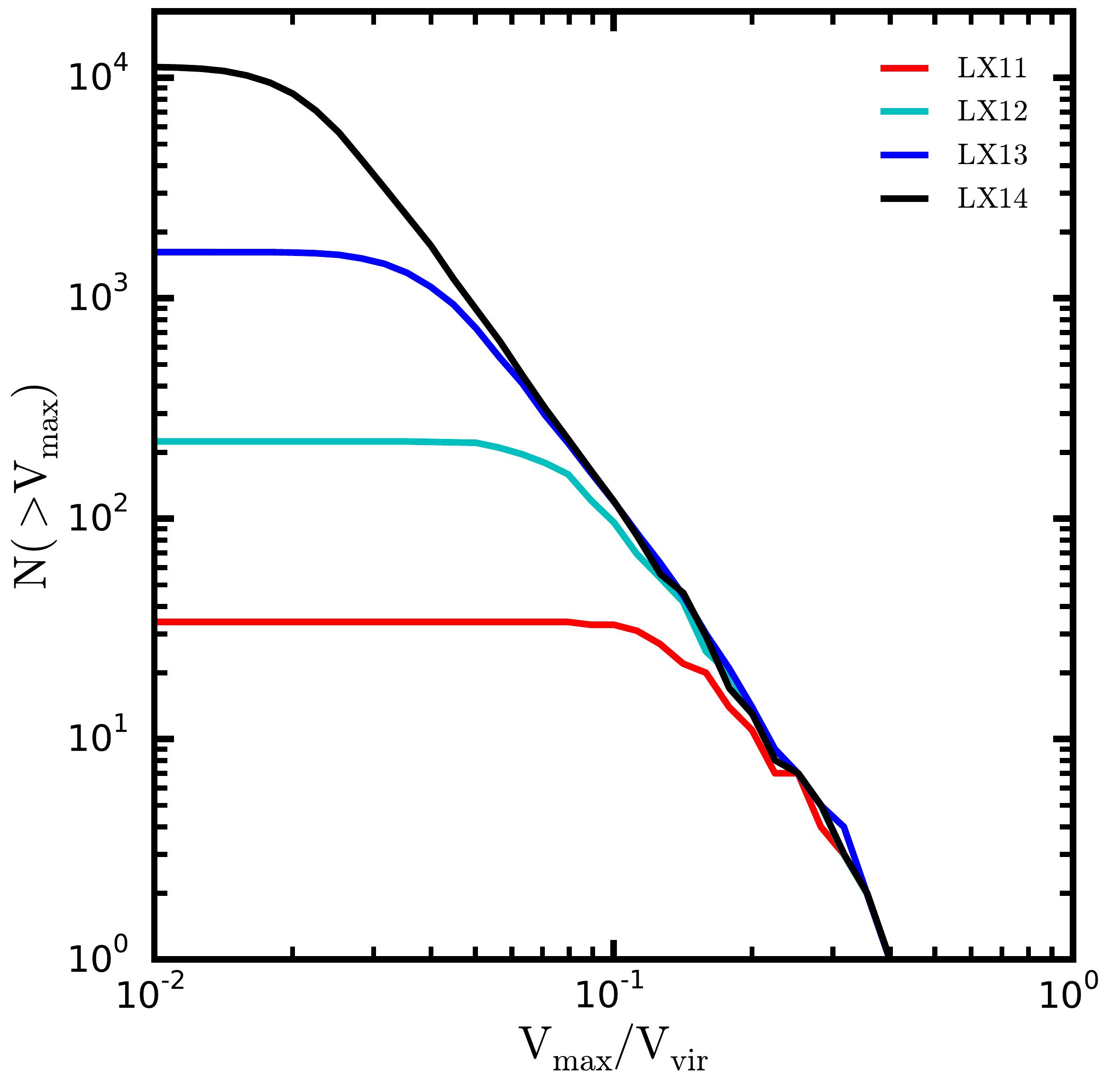} 
\caption{Convergence test for the Cat-2 and Cat-9 halos illustrating convergent halo profiles and \Vmax functions respectively. Panel (a) normalized halo profiles where the thick line represents the density profile above the radius defined by \cite{Power03}. The thin dashed line of the same color is the profile extrapolated down to the softening length. Panel (b) the cumulative abundance of subhalos as a function of their maximum circular velocity (\Vmax) . These halos are representative of the convergence we achieve for all $relaxed$ $Caterpillar$ halos.}
\label{fig:convergence}
\end{figure}

In Figure \ref{fig:convergence} we plot the halo profiles (Cat-2) and maximum circular velocity functions (Cat-9) at all our resolutions. We find our halos are well converged down to $\sim$0.2$\%$ of \Rvir. In the case of the \Vmax functions, we find we are converged down to $\sim$4 \kms at our highest resolution. When normalized to the host halo virial velocity, the halos are in excellent agreement with one another. Halos were re-simulated at successively higher and higher mass and spatial resolution from the initial parent volume. In each instance, care was taken to ensure all halo properties were numerically converged (provided that quantity was not resolution limited, e.g.\ halo shape). In Table \ref{tab:iccontam1} and \ref{tab:iccontam2} we show the same quantities as in Table \ref{tab:catsuite} from the text but this time include the lower resolution halo properties. A full resolution study will be provided at the website, \url{http://www.caterpillarproject.org} when the {\sc{LX15}} runs have been completed.

\begin{center}
\begin{table*}[!h]
\caption{Halo properties for the first set of 12 $Caterpillar$ halos at each resolution.}
\label{tab:iccontam1}
\begin{minipage}{1805mm}
\begin{tabular}{ccccccccccccccc}
 \hline
Name & Geometry & $n$\Rvir & {\sc{lx}} & \Mvir & \Rvir & c & \Vmax & \Rmax & $z_{0.5}$ & $z_{lmm}$  & $f_{\mathrm{m,subs}}$ &  b/c & c/a & $R_{\mathrm{res}}$ \\
  &  &  &  &  ($10^{12}$ \Msol) &  (kpc) &   & (\kms) &  (kpc) & & & & & & (Mpc) \\
  \hline
Cat-1 & EA & 5 & 11 & 1.579 & 307.690 & 7.762 & 172.293 & 35.049 & 0.881 & 2.118 & 0.161 & 0.810 & 0.828 & 1.391 \\
 &  &  & 12 & 1.560 & 306.491 & 7.494 & 171.060 & 34.292 & 0.894 & 2.118 & 0.171 & 0.824 & 0.869 & 1.248 \\
 &  &  & 13 & 1.560 & 306.458 & 7.647 & 170.707 & 36.451 & 0.894 & 2.157 & 0.197 & 0.842 & 0.883 & 1.138 \\
 &  &  & 14 & 1.559 & 306.378 & 7.492 & 169.756 & 34.083 & 0.894 & 2.157 & 0.207 & 0.841 & 0.869 & 0.998 \\
 \hline
Cat-2 & EB & 4 & 11 & 1.807 & 321.876 & 7.621 & 176.924 & 64.931 & 0.742 & 0.719 & 0.092 & 0.596 & 0.724 & 1.577 \\
 &  &  & 12 & 1.782 & 320.357 & 8.575 & 179.069 & 53.856 & 0.742 & 0.719 & 0.112 & 0.607 & 0.716 & 1.522 \\
 &  &  & 13 & 1.792 & 320.970 & 8.382 & 178.753 & 54.360 & 0.742 & 0.731 & 0.137 & 0.643 & 0.731 & 1.480 \\
 &  &  & 14 & 1.791 & 320.907 & 8.374 & 178.851 & 55.268 & 0.742 & 0.731 & 0.148 & 0.636 & 0.719 & 1.463 \\
 \hline
Cat-3 & EB & 4 & 11 & 1.343 & 291.538 & 10.763 & 175.066 & 26.554 & 0.802 & 0.790 & 0.079 & 0.961 & 0.971 & 1.966 \\
 &  &  & 12 & 1.355 & 292.387 & 10.489 & 175.142 & 29.083 & 0.802 & 0.790 & 0.100 & 0.868 & 0.915 & 1.926 \\
 &  &  & 13 & 1.355 & 292.400 & 10.523 & 172.946 & 31.565 & 0.802 & 0.802 & 0.117 & 0.850 & 0.905 & 1.906 \\
 &  &  & 14 & 1.354 & 292.300 & 10.170 & 172.440 & 31.701 & 0.802 & 0.802 & 0.136 & 0.865 & 0.927 & 1.894 \\
 \hline
Cat-4 & EB & 4 & 11 & 1.503 & 302.676 & 8.308 & 169.309 & 91.132 & 0.894 & 0.908 & 0.128 & 0.749 & 0.825 & 1.791 \\
 &  &  & 12 & 1.415 & 296.632 & 9.208 & 168.170 & 85.989 & 0.922 & 0.922 & 0.120 & 0.681 & 0.762 & 1.594 \\
 &  &  & 13 & 1.434 & 298.009 & 8.434 & 164.999 & 59.225 & 0.936 & 0.922 & 0.156 & 0.673 & 0.743 & 1.561 \\
 &  &  & 14 & 1.424 & 297.295 & 8.573 & 164.344 & 53.466 & 0.936 & 0.922 & 0.175 & 0.671 & 0.739 & 1.531 \\
 \hline
Cat-5 & EB & 4 & 11 & 1.306 & 288.846 & 11.897 & 173.913 & 33.844 & 0.584 & 0.519 & 0.025 & 0.551 & 0.835 & 1.676 \\
 &  &  & 12 & 1.318 & 289.714 & 11.896 & 174.223 & 36.356 & 0.574 & 0.519 & 0.041 & 0.547 & 0.765 & 1.657 \\
 &  &  & 13 & 1.314 & 289.450 & 12.324 & 176.818 & 29.346 & 0.574 & 0.519 & 0.055 & 0.556 & 0.825 & 1.617 \\
 &  &  & 14 & 1.309 & 289.079 & 12.108 & 176.399 & 32.103 & 0.564 & 0.510 & 0.069 & 0.552 & 0.815 & 1.608 \\
 \hline
Cat-6 & EB & 4 & 11 & 1.371 & 293.516 & 10.373 & 172.100 & 32.944 & 1.144 & 1.275 & 0.094 & 0.495 & 0.534 & 1.708 \\
 &  &  & 12 & 1.347 & 291.848 & 10.522 & 172.873 & 30.622 & 1.161 & 1.275 & 0.116 & 0.496 & 0.525 & 1.495 \\
 &  &  & 13 & 1.366 & 293.186 & 10.086 & 170.858 & 32.794 & 1.161 & 1.295 & 0.138 & 0.510 & 0.529 & 1.300 \\
 &  &  & 14 & 1.363 & 292.946 & 10.196 & 171.647 & 33.632 & 1.161 & 1.295 & 0.153 & 0.508 & 0.528 & 1.295 \\
 \hline
Cat-7 & EB & 4 & 11 & 1.142 & 276.168 & 2.513 & 139.055 & 140.859 & 0.065 & 0.057 & 0.693 & 0.191 & 0.301 & 1.756 \\
 &  &  & 12 & 1.111 & 273.686 & 2.487 & 136.803 & 145.085 & 0.074 & 0.057 & 0.615 & 0.170 & 0.288 & 1.520 \\
 &  &  & 13 & 1.091 & 272.009 & 1.674 & 133.574 & 162.291 & 0.065 & 0.036 & 0.693 & 0.168 & 0.235 & 1.510 \\
 &  &  & 14 & 1.092 & 272.099 & 1.757 & 134.148 & 157.438 & 0.070 & 0.032 & 0.735 & 0.151 & 0.207 & 1.477 \\
 \hline
Cat-8 & EB & 4 & 11 & 1.729 & 317.150 & 13.081 & 198.577 & 46.800 & 1.541 & 2.195 & 0.032 & 0.602 & 0.768 & 1.690 \\
 &  &  & 12 & 1.716 & 316.337 & 13.154 & 198.229 & 39.671 & 1.315 & 2.195 & 0.053 & 0.594 & 0.775 & 1.597 \\
 &  &  & 13 & 1.701 & 315.450 & 13.340 & 197.637 & 39.810 & 1.516 & 2.235 & 0.066 & 0.599 & 0.791 & 1.550 \\
 &  &  & 14 & 1.702 & 315.466 & 13.507 & 198.564 & 40.819 & 1.516 & 2.235 & 0.078 & 0.605 & 0.787 & 1.540 \\
 \hline
Cat-9 & EB & 4 & 11 & 1.330 & 290.616 & 12.568 & 177.522 & 32.309 & 1.236 & 1.217 & 0.050 & 0.493 & 0.762 & 2.383 \\
 &  &  & 12 & 1.331 & 290.654 & 11.616 & 175.047 & 27.903 & 1.236 & 1.236 & 0.070 & 0.486 & 0.754 & 2.101 \\
 &  &  & 13 & 1.329 & 290.538 & 12.132 & 176.808 & 30.297 & 1.255 & 1.236 & 0.085 & 0.500 & 0.754 & 1.833 \\
 &  &  & 14 & 1.322 & 289.987 & 12.401 & 177.414 & 30.336 & 1.255 & 1.236 & 0.094 & 0.513 & 0.762 & 2.080 \\
 \hline
Cat-10 & EB & 4 & 11 & 1.319 & 289.809 & 11.902 & 175.553 & 41.894 & 1.699 & 2.010 & 0.052 & 0.561 & 0.709 & 1.983 \\
 &  &  & 12 & 1.332 & 290.764 & 11.439 & 174.479 & 29.806 & 1.516 & 2.010 & 0.069 & 0.551 & 0.679 & 1.870 \\
 &  &  & 13 & 1.328 & 290.477 & 11.714 & 175.124 & 25.839 & 1.644 & 2.010 & 0.088 & 0.559 & 0.703 & 1.740 \\
 &  &  & 14 & 1.323 & 290.119 & 11.714 & 174.989 & 39.721 & 1.644 & 2.010 & 0.103 & 0.559 & 0.703 & 1.775 \\
 \hline
Cat-11 & EB & 4 & 11 & 1.194 & 280.361 & 10.551 & 165.980 & 62.881 & 1.059 & 4.368 & 0.175 & 0.527 & 0.719 & 1.490 \\
 &  &  & 12 & 1.196 & 280.471 & 10.044 & 163.290 & 70.202 & 1.043 & 4.368 & 0.200 & 0.525 & 0.703 & 1.408 \\
 &  &  & 13 & 1.190 & 280.043 & 12.272 & 173.893 & 45.727 & 1.059 & 1.644 & 0.199 & 0.590 & 0.868 & 1.192 \\
 &  &  & 14 & 1.179 & 279.187 & 12.522 & 172.723 & 53.187 & 1.059 & 4.368 & 0.215 & 0.597 & 0.867 & 1.135 \\
 \hline
Cat-12 & EA & 5 & 11 & 1.786 & 320.627 & 11.723 & 191.564 & 59.256 & 1.336 & 2.542 & 0.034 & 0.592 & 0.724 & 1.664 \\
 &  &  & 12 & 1.749 & 318.388 & 11.824 & 192.085 & 56.859 & 1.336 & 2.542 & 0.042 & 0.572 & 0.686 & 1.342 \\
 &  &  & 13 & 1.767 & 319.441 & 11.663 & 191.320 & 49.435 & 1.336 & 2.542 & 0.062 & 0.571 & 0.703 & 1.239 \\
 &  &  & 14 & 1.763 & 319.209 & 11.402 & 191.259 & 52.717 & 1.336 & 9.616 & 0.073 & 0.584 & 0.645 & 1.162 \\
 \hline
\end{tabular}

{Notes: The resolution details for each refinement level (i.e.\ 11, 12, 13, 14) can be found in Table \ref{tab:sim} and the geometry definitions in Table \ref{tab:icgeom}.}
\end{minipage}
\end{table*}
\end{center}

\newpage
\begin{center}
\begin{table*}
\caption{Halo properties for the second set of 12 $Caterpillar$ halos at each resolution.}
\label{tab:iccontam2}
\begin{minipage}{1805mm}
\begin{tabular}{ccccccccccccccc}
 \hline
Name & Geometry & $n$\Rvir & {\sc{lx}} & \Mvir & \Rvir & c & \Vmax & \Rmax & $z_{0.5}$ & $z_{lmm}$   & $f_{\mathrm{m,subs}}$ &  b/c & c/a & $R_{\mathrm{res}}$ \\
  &  &  &  &  ($10^{12}$\Msol) &  (kpc) &   & (km/s) &  (kpc) & & & &  & & (Mpc) \\
  \hline
Cat-13 & EB & 4 & 11 & 1.168 & 278.303 & 12.664 & 169.603 & 31.214 & 1.180 & 11.092 & 0.042 & 0.595 & 0.652 & 2.069 \\
 &  &  & 12 & 1.171 & 278.509 & 12.979 & 170.750 & 31.408 & 1.161 & 14.748 & 0.063 & 0.575 & 0.634 & 1.742 \\
 &  &  & 13 & 1.163 & 277.896 & 13.052 & 171.892 & 34.163 & 1.161 & 15.750 & 0.073 & 0.580 & 0.655 & 1.634 \\
 &  &  & 14 & 1.164 & 277.938 & 12.850 & 171.222 & 33.757 & 1.161 & 11.092 & 0.090 & 0.578 & 0.645 & 1.566 \\
 \hline
Cat-14 & EC & 4 & 11 & 0.744 & 239.430 & 9.526 & 137.580 & 42.772 & 1.180 & 4.155 & 0.060 & 0.714 & 0.851 & 2.516 \\
 &  &  & 12 & 0.757 & 240.865 & 8.854 & 136.512 & 27.875 & 1.144 & 4.258 & 0.086 & 0.709 & 0.849 & 2.301 \\
 &  &  & 13 & 0.754 & 240.529 & 9.148 & 137.266 & 44.395 & 1.144 & 4.258 & 0.097 & 0.694 & 0.842 & 2.234 \\
 &  &  & 14 & 0.750 & 240.119 & 9.135 & 137.437 & 26.660 & 1.144 & 4.258 & 0.113 & 0.705 & 0.859 & 2.178 \\
 \hline
Cat-15 & EX & 5 & 11 & 1.501 & 302.562 & 8.950 & 173.834 & 31.210 & 1.144 & 3.165 & 0.072 & 0.897 & 0.912 & 1.669 \\
 &  &  & 12 & 1.497 & 302.281 & 9.223 & 174.792 & 33.832 & 1.144 & 3.165 & 0.089 & 0.897 & 0.926 & 1.630 \\
 &  &  & 13 & 1.504 & 302.755 & 9.077 & 174.431 & 36.520 & 1.144 & 3.165 & 0.111 & 0.837 & 0.861 & 1.597 \\
 &  &  & 14 & 1.505 & 302.787 & 8.983 & 174.124 & 37.043 & 1.144 & 3.165 & 0.126 & 0.849 & 0.877 & 1.119 \\
 \hline
Cat-16 & EB & 4 & 11 & 0.993 & 263.614 & 10.997 & 154.748 & 42.280 & 1.315 & 3.165 & 0.053 & 0.567 & 0.765 & 1.406 \\
 &  &  & 12 & 0.976 & 262.082 & 12.099 & 156.589 & 28.820 & 1.315 & 3.165 & 0.072 & 0.593 & 0.791 & 1.393 \\
 &  &  & 13 & 0.980 & 262.447 & 11.888 & 156.193 & 29.498 & 1.315 & 3.165 & 0.088 & 0.597 & 0.766 & 1.384 \\
 &  &  & 14 & 0.982 & 262.608 & 11.737 & 155.362 & 28.768 & 1.315 & 3.165 & 0.106 & 0.618 & 0.792 & 0.671 \\
  \hline
 Cat-17 & EX & 4 & 11 & 1.311 & 289.204 & 13.216 & 178.671 & 38.818 & 1.846 & 1.943 & 0.038 & 0.646 & 0.794 & 1.525 \\
 &  &  & 12 & 1.314 & 289.456 & 12.906 & 178.676 & 39.713 & 1.846 & 1.943 & 0.057 & 0.680 & 0.875 & 1.427 \\
 &  &  & 13 & 1.329 & 290.487 & 12.505 & 178.763 & 38.717 & 1.846 & 1.976 & 0.084 & 0.657 & 0.863 & 1.333 \\
 &  &  & 14 & 1.319 & 289.800 & 12.765 & 179.056 & 38.329 & 1.846 & 1.976 & 0.093 & 0.664 & 0.881 & 1.299 \\
 \hline
Cat-18 & EX & 4 & 11 & 1.428 & 297.536 & 7.909 & 167.184 & 32.058 & 0.451 & 0.427 & 0.100 & 0.677 & 0.847 & 1.491 \\
 &  &  & 12 & 1.414 & 296.559 & 7.861 & 164.702 & 48.041 & 0.459 & 0.412 & 0.123 & 0.720 & 0.840 & 1.397 \\
 &  &  & 13 & 1.400 & 295.596 & 7.823 & 165.164 & 40.766 & 0.493 & 0.435 & 0.141 & 0.622 & 0.712 & 1.228 \\
 &  &  & 14 & 1.407 & 296.099 & 7.887 & 163.920 & 57.217 & 0.493 & 0.435 & 0.159 & 0.676 & 0.816 & 0.397 \\
 \hline
Cat-19 & EX & 5 & 11 & 1.179 & 279.143 & 10.467 & 164.816 & 34.292 & 1.566 & 2.693 & 0.113 & 0.640 & 0.857 & 1.933 \\
 &  &  & 12 & 1.174 & 278.788 & 10.158 & 163.679 & 34.514 & 1.566 & 2.693 & 0.132 & 0.668 & 0.919 & 1.861 \\
 &  &  & 13 & 1.177 & 279.002 & 10.139 & 163.868 & 30.433 & 1.541 & 2.118 & 0.149 & 0.672 & 0.933 & 1.800 \\
 &  &  & 14 & 1.174 & 278.770 & 10.468 & 164.726 & 29.112 & 1.541 & 2.118 & 0.169 & 0.664 & 0.937 & 1.712 \\
 \hline
Cat-20 & BB & 4 & 11 & 0.765 & 241.720 & 13.409 & 150.030 & 25.189 & 1.516 & 5.588 & 0.045 & 0.608 & 0.743 & 1.677 \\
 &  &  & 12 & 0.756 & 240.683 & 13.443 & 148.881 & 27.312 & 1.541 & 5.588 & 0.053 & 0.634 & 0.775 & 1.521 \\
 &  &  & 13 & 0.761 & 241.208 & 13.456 & 149.682 & 29.340 & 1.516 & 5.761 & 0.084 & 0.613 & 0.752 & 1.377 \\
 &  &  & 14 & 0.763 & 241.484 & 13.324 & 149.672 & 30.417 & 1.492 & 5.427 & 0.099 & 0.601 & 0.733 & 1.311 \\
 \hline
Cat-21 & EX & 4 & 11 & 1.865 & 325.250 & 11.820 & 193.253 & 42.842 & 1.144 & 1.198 & 0.042 & 0.456 & 0.584 & 1.551 \\
 &  &  & 12 & 1.876 & 325.890 & 10.950 & 191.015 & 54.116 & 1.109 & 1.161 & 0.075 & 0.475 & 0.637 & 1.426 \\
 &  &  & 13 & 1.889 & 326.663 & 10.465 & 189.607 & 57.507 & 1.126 & 1.198 & 0.103 & 0.472 & 0.590 & 1.342 \\
 &  &  & 14 & 1.881 & 326.206 & 10.618 & 190.683 & 50.954 & 1.126 & 1.198 & 0.118 & 0.482 & 0.611 & 1.453 \\
 \hline
Cat-22 & EX & 5 & 11 & 1.560 & 306.489 & 9.356 & 177.811 & 33.807 & 0.828 & 5.940 & 0.044 & 0.496 & 0.643 & 2.003 \\
 &  &  & 12 & 1.594 & 308.677 & 9.799 & 181.703 & 44.919 & 0.790 & 5.940 & 0.052 & 0.461 & 0.637 & 1.903 \\
 &  &  & 13 & 1.497 & 302.257 & 10.655 & 180.773 & 37.743 & 0.854 & 5.940 & 0.068 & 0.518 & 0.695 & 1.837 \\
 &  &  & 14 & 1.495 & 302.116 & 10.666 & 180.647 & 35.860 & 0.841 & 29.488 & 0.080 & 0.512 & 0.694 & 1.744 \\
 \hline
Cat-23 & EX & 4 & 11 & 1.608 & 309.596 & 11.989 & 189.267 & 33.023 & 1.180 & 10.062 & 0.051 & 0.635 & 0.845 & 1.623 \\
 &  &  & 12 & 1.604 & 309.328 & 12.865 & 191.457 & 32.232 & 1.180 & 9.616 & 0.071 & 0.589 & 0.729 & 1.236 \\
 &  &  & 13 & 1.613 & 309.926 & 12.135 & 190.191 & 31.524 & 1.161 & 9.616 & 0.080 & 0.602 & 0.763 & 1.245 \\
 &  &  & 14 & 1.607 & 309.524 & 12.489 & 190.705 & 32.421 & 1.161 & 9.616 & 0.094 & 0.607 & 0.784 & 1.207 \\
 \hline
Cat-24 & EB & 4 & 11 & 1.329 & 290.537 & 11.152 & 174.259 & 43.136 & 1.217 & 2.801 & 0.038 & 0.651 & 0.705 & 1.260 \\
 &  &  & 12 & 1.323 & 290.054 & 11.326 & 175.088 & 48.435 & 1.217 & 2.801 & 0.052 & 0.645 & 0.674 & 1.396 \\
 &  &  & 13 & 1.335 & 290.969 & 11.490 & 177.313 & 34.438 & 1.144 & 2.801 & 0.077 & 0.675 & 0.721 & 1.190 \\
 &  &  & 14 & 1.334 & 290.866 & 11.378 & 176.911 & 36.800 & 1.144 & 3.608 & 0.090 & 0.689 & 0.734 & 1.102 \\
\hline
\end{tabular}

{Notes: The resolution details for each refinement level (i.e.\ 11, 12, 13, 14) can be found in Table \ref{tab:sim} and the geometry definitions in Table \ref{tab:icgeom}.}
\end{minipage}
\end{table*}
\end{center}
\end{document}